\documentclass{emulateapj}

\usepackage{amsmath}
\usepackage{amssymb}
\usepackage{amsfonts}
\usepackage{amstext}
\usepackage{graphicx}
\usepackage{fancybox}
\usepackage[usenames,dvipsnames]{color}
\usepackage{pstricks}
\usepackage{bm}
\usepackage{soul, color}
\usepackage{latexsym}
\usepackage{pifont}
\usepackage{shorttoc}
\usepackage{subfigure}
\usepackage{textcomp} 
\usepackage{hyperref}
\usepackage{bm}
\usepackage{lipsum}
\usepackage{threeparttable}
\usepackage{enumitem}
\def\aap{A\&A}

\def\apjs{ApJS}
\def\apjl{ApJL}

\newcommand{\no}[1]{}
\renewcommand{\exp}[1]{\mathrm{exp}\left(#1\right)}
\def\gsim{\mathrel{\rlap{\lower4pt\hbox{\hskip1pt$\sim$}}
    \raise1pt\hbox{$>$}}}                

\newcommand{\myemail}{l.m.ribas@astro.uio.no}
\def\lsim{~\rlap{$<$}{\lower 1.0ex\hbox{$\sim$}}}

\def\gsim{~\rlap{$>$}{\lower 1.0ex\hbox{$\sim$}}}

\shorttitle{The contribution of fluorescence to LAHs}
\shortauthors{Llu\'is Mas-Ribas \& Mark Dijkstra }

\begin{document}

\title{On the contribution of fluorescence to lyman alpha halos \\
around star forming galaxies }

\author{Llu\'is Mas-Ribas\altaffilmark{1} and Mark Dijkstra}
\affil{Institute of Theoretical Astrophysics, University of Oslo,
P.O. Box 1029 Blindern, N-0315 Oslo, Norway}

\altaffiltext{1}{\myemail}

\begin{abstract}
We quantify the contribution of Ly$\alpha$ fluorescence to observed spatially extended 
Ly$\alpha$ halos around Ly$\alpha$ emitters (LAE) at redshift ${\rm z=3.1}$. The key physical quantities that
describe the fluorescent signal include ({\it i}) the distribution of cold gas in the circum-galactic
medium (CGM); we explore simple analytic models and fitting functions to recent 
hydrodynamical simulations; ({\it ii}) local variations in the ionizing background due
to ionizing sources that cluster around the central galaxy. We account  for clustering by boosting the observationally inferred volumetric production rate of ionizing photons, $\epsilon_{\rm LyC}$, by a factor $1+\xi_{\rm LyC}(r)$, in which $\xi_{\rm LyC}(r)$ quantifies the clustering of ionizing sources around the central galaxy. We compute $\xi_{\rm LyC}(r)$ by assigning an ``effective'' bias parameter to the ionizing sources.  This novel approach allows us to quantify our ignorance of the population of ionizing sources in a simple parametrized form. We find a maximum enhancement 
in the local ionizing background in the range $50-200$ at $r \sim 10$ physical kpc. For spatially 
uncorrelated ionizing sources and fluorescing clouds we find that fluorescence can contribute 
up to $\sim 50-60\%$ of the observed spatially extended Ly$\alpha$ emission. We briefly discuss 
how future 
observations can shed light on the nature of Ly$\alpha$ halos around star forming galaxies.
\end{abstract}



\section{Introduction}\label{sec:intro}

  The gaseous region out to $300\, {\rm kpc}$ around galaxies, the so called 
circum-galactic medium (CGM), plays an important
role when it comes to studying the properties and characteristics of the galaxy population 
\citep[e.g.,][]{Bahcall1969}. The CGM connects galaxies and the intergalactic medium (IGM), via
gas accretion inflows and feedback-driven outflows.
Therefore, studying these regions at different redshifts provides invaluable 
information about the properties, formation and evolution
of galaxies \citep[e.g.,][]{Steidel2010,Steidel2011,Tumlinson2011,DijkstraKramer2012, 
Cantalupo2012, Cantalupo2014, Rahmati2015}. 

The Lyman alpha (Ly$\alpha$ hereafter) radiation, which corresponds to the 
transition from the $2^{\,2}\,\rm P$ state to the 
ground state of the hydrogen atom with a line centre wavelength at $1215.67\,{\rm \AA}$, 
provides an excellent tool for studying hydrogen gas in the CGM
\citep[see][for detailed reviews]{Barnes2014,Dijkstra2014,Hayes2015}.
The CGM has been detected in Ly$\alpha$ absorption using galaxy-galaxy 
\citep[e.g.,][]{Rudie2012}, galaxy-quasar
\citep[e.g.,][]{Rakic2012} and quasar-quasar \citep[e.g.,][]{HennawiProchaska2013, 
Prochaska2013} pairs. However, Ly$\alpha$ has also been observed in emission 
around low redshift galaxies \citep[e.g.,][]{Ostlin2009,Guaita2015},
radio-loud galaxies \citep[RLG; e.g.,][]{Saito2015} and Ly$\alpha$ blobs \citep[e.g.,][]
{Fynbo1999,Steidel2000,Matsuda2004, Saito2006,Ao2015}. 
In addition, recent analyses have indicated that all star forming galaxies may be surrounded 
by a faint Ly$\alpha$ emission from the CGM, at surface brightness levels 
of $\rm{\sim 10^{-19} erg \,s^{-1}\, cm^{-2}\, arcsec^{-2}}$. Some studies used 
stacking analyses to uncover this faint signal out to $\sim100\,{\rm physical\, kpc}$ 
\citep[e.g.,][]{Hayashino2004, Steidel2011,Matsuda2012,Momose2014, Momose2015}, 
although not all obtained positive detections \citep{Jiang2013,Feldmeier2013}. 
Individual detections have also been reported, and some of them account for the  
inclusion of the spectra \citep[e.g.,][see also \cite{Rauch2008a, Rauch2008b}]{Wisotzki2015}.
The Ly$\alpha$ spectral line shape contains information on the gas kinematics
\citep{Shapley2003,Verhamme2006, DijkstraKramer2012}, while the spatial extent of 
Ly$\alpha$ emission (and absorption) traces the spatial distribution of hydrogen gas
\citep{Steidel2010,Steidel2011}.

The origin of the spatially extended Ly$\alpha$ emission around galaxies 
(Ly$\alpha$ halos; LAHs)
can be attributed to ({\it i}) scattering: neutral hydrogen gas in the CGM can {\it scatter}
Ly$\alpha$ photons that were produced in the inner part of galaxies, and which 
subsequently escaped from the interstellar medium 
\citep[see, e.g.,][]{Laursen2007,Laursen2011,Zheng2010,Steidel2011, Zheng2011, 
DijkstraKramer2012, Verhamme2012}; and ({\it ii}) \textit{in situ} production: Ly$\alpha$ can be 
emitted by hydrogen atoms in the CGM following collisional excitation and/or recombination. 
Collisional 
excitation of the Ly$\alpha$ transition in atomic hydrogen by free electrons   
reduces the thermal energy - and thus cools - the circum-galactic gas 
 \citep[e.g.,][]{Haiman2000, Fardal2001, DijkstraLoeb2009, FaucherGiguere2010,Rosdahl2012}. 
Radiative ionizations can produce Ly$\alpha$ photons in a process called
`fluorescence': ionizing photons that are produced inside a galaxy  \citep{Furlanetto2005, 
Weidinger2005,Nagamine2010}, but also those photons coming from distant sources such 
as quasars or other galaxies \citep{Haiman2001,Alam2002, Cantalupo2005, Adelberger2006, Cantalupo2007, 
Rauch2008a, Rauch2008b,Kollmeier2010,HennawiProchaska2013, Cantalupo2014}, can 
photoionize the circum-galactic hydrogen which recombines almost immediately. Approximately 
$66\%$ ($42\%$) of the recombinations\footnote{Note that absorption of higher-order Lyman 
series photons can also give rise to fluorescent Ly$\alpha$ emission. This process can increase 
Ly$\alpha$ emission as discussed in, e.g., \cite{Furlanetto2005}.} produce Ly$\alpha$ for case-B (case-A) 
recombination \citep{GouldWeinberg1996}.
It is currently still unclear what the main physical mechanism behind LAHs is, and what their relation 
with the more luminous Ly$\alpha$ `blobs' is.

In this work, we revisit the effect of fluorescence. This is motivated by the observed dependence 
of Ly$\alpha$ surface brightness profiles on the local overdensity: as the mean number 
of Ly$\alpha$ 
emitting galaxies within a sphere of radius $\sim2$ Mpc h$^{-1}$ increases, the spatial extend of 
the Ly$\alpha$ halos increases \citep{Steidel2011, Matsuda2012, Chiang2015}. This dependence 
is naturally expected when the Ly$\alpha$ emission is powered by fluorescence, as we expect the 
(average) intensity of the ionizing radiation field to be enhanced in overdense regions. Our main 
goal is to investigate whether, and under which conditions, fluorescence in the CGM can give rise 
to the observed Ly$\alpha$ halos (and their dependence on overdensity). This is an interesting 
question to address because, as we will show in this work, it constrains spatial variations in the 
ionizing background, which in turn relate to the escape of ionizing photons from galaxies. Our 
work focuses on Ly$\alpha$ emitters (LAEs) at redshift $z=3.1$. This allows us to compare with the work by 
\cite{Matsuda2012}, who presented their results with respect to LAE overdensity at this 
redshift. At z$=3.1$ stellar radiation provides the major contribution to 
the ionizing background \citep[e.g.,][]{FaucherGiguere2008}. In a companion paper, we 
will focus on the contribution of Ly$\alpha$ emission from satellite galaxies to LAHs.

The outline of this paper is as follows: we present our main formalism 
for computing the surface brightness in fluorescent Ly$\alpha$ in \S~\ref {sec:formalism}.
This formalism shows that the two key model ingredients are (\textit{i}) the local 
photoionization rate $\Gamma(r)$, and (\textit{ii}) the distribution of neutral hydrogen gas in
the CGM. We present the models for the calculation of these ingredients in \S~\ref{sec:model}, 
and we show our main results in \S~\ref{sec:results}. We discuss the shortcomings of our 
model in \S~\ref{sec:discussion}, before concluding in \S~\ref{sec:conclusion}.


\section{General formalism}
\label{sec:formalism}

In this section, we introduce the general formalism for the calculation of the Ly$\alpha$ surface 
brightness profile. This will highlight the main ingredients that have to be assessed in 
more detail, namely ({\it i}) the characteristics of the circum-galactic medium around the central 
galaxy, and ({\it ii}) the ionizing radiation field affecting this medium.

It is well known that the environment of galaxies contains cold optically thick gas \citep[e.g.,][]
{Bosma1981,Bergeron1986,Kunth1998, Momiralda1996,Steidel2010}.
We assume that the central galaxy contains a spherically symmetric distribution of cold gas 
clumps embedded within a hot medium. It is important to mention here that our
formalism does not take into account a correlation between self shielding gas and the surrounding
ionizing sources, thus making our calculations conservative. We briefly discuss the importance
of this correlation for the boost of the fluorescent signal in \S~\ref{sec:discussion}.
The distribution of clumps can be characterized by their `covering factor', $f_c(r)$, which denotes 
the number of self-shielding clumps along a differential length at a distance $r$ from the central 
galaxy (which in the present paper we consider to be an LAE). Many useful quantities can be expressed in terms of $f_c(r)$.

For example, the total number of clumps along a sightline at impact parameter $b$ is
$N_{\rm clump}(b) = \int_{-\infty}^{\infty} {\rm d}s\,f_c(r[b,s])$, 
where $s$ denotes the line-of-sight coordinate. The distance from the galaxy is $r=\sqrt{b^2+s^2}$, 
and so we can write 
\begin{equation}\label{eq:nclump}
N_{\rm clump}(b) = 2\,\int_b^{\infty} \frac{r\,{\rm d}r}{\sqrt{r^2\,-\,b^2}}f_c(r)~.
\end{equation} The parameter $N_{\rm clump}(b)$ denotes the {\it average} number of clumps, 
and the actual number at a given impact parameter fluctuates around this mean due to the discrete 
nature of the clumps. The probability 
of finding {\it at least} one clump at impact parameter $b$ is $P(N\geq1)=1-P(N=0)$, where $P(N=0)$ 
denotes the Poisson probability of finding no clumps at impact parameter $b$. 
For a Poisson distribution, 
the probability of finding at least one clump at impact parameter $b$ is therefore
\begin{equation}\label{eq:Nclos}
P(N_{\rm clump} \geq 1, b) = 1 - \exp{- N_{\rm clump}(b)} ~.
\end{equation} 
The quantity 
$P(N_{\rm clump} \geq 1, b)$ is useful when comparing to recent results from hydrodynamical simulations.

We can also compute the Ly$\alpha$ surface brightness produced by fluorescence at a given impact 
parameter $b$ from $f_{\rm c}(r)$ as 
\begin{equation}
SB_{\rm Ly\alpha}(b) = \int_{-\infty}^{\infty}{\rm d}s\,P_{\rm clump}(b,s)\,SB_{\rm Ly\alpha}^{clump}(b,s)~,
\end{equation} where $P_{\rm clump}(b,s){\rm d}s = f_c(b,s){\rm d}s=f_c(r){\rm d}s$ denotes the differential probability 
of finding a clump along within the range $s \pm {\rm d}s/2$, in which $s$ denotes the line-of-sight coordinate.
Furthermore, $SB_{\rm Ly\alpha}^{\rm clump}(b,s)= SB_{\rm Ly\alpha}^{\rm clump}(r) = SB_{\rm Ly\alpha}
^{\rm HM}\times\frac{\Gamma_{\rm{HI}}(r)}{\Gamma_{\rm{HI}}^{\rm HM}}$. Here $SB_{\rm Ly\alpha} ^{\rm HM} 
=3.67\times10^{-20}[(1\,+\,z)/4]^{-4}$ ${\rm erg\, s^{-1} \, cm^{-2}\, arcsec^{-2}}$, 
is the value obtained for fluorescence by \cite{Cantalupo2005} for a photoionization rate 
$\Gamma_{\rm{HI}}^{\rm HM}=1.15 \times 10^{-12}\,{\rm s^{-1}}$. Rewriting as above we obtain
\begin{equation}\label{eq:sb}
SB_{\rm Ly\alpha}(b) = \frac{2\,SB_{\rm Ly\alpha} ^{\rm HM}}{\Gamma_{\rm{HI}}^{
\rm HM}}\int_{b}^{\infty}\frac{r\,{\rm d}r}{\sqrt{r^2\,-\,b^2}}\,f_c(r)\,\Gamma_{\rm{HI}}(r) ~,
\end{equation} where we emphazise that $\Gamma_{\rm{HI}}(r)$ denotes the photoionization rate at 
a distance $r$ from the central galaxy. This `local' photoionization rate can differ significantly from the 
overall background that is inferred from, e.g., Ly$\alpha$ forest studies.  

 Eq.~\ref{eq:sb} highlights two key ingredients that determine the Ly$\alpha$ 
surface brightness. These include:
\begin{enumerate}
\item The covering factor, $f_{\rm c}(r)$, which describes the spatial distribution of self-shielding 
clouds in the CGM.
\item The value of the local photoionization rate, $\Gamma_{\rm{HI}}(r)$, which is determined by ionizing 
	sources surrounding the galaxy, as well as the ionizing luminosity from the central galaxy.
\end{enumerate}

It is worth stressing that both these ingredients are difficult to model from first principles: (\textit{i}) the distribution of cold gas in the CGM depends on the adopted feedback prescriptions 
\citep[e.g.,][]{FaucherGiguere2015,FaucherGiguere2016}. Recent observations indicate that there is a substantial amount of cold gas locked up in small ($\sim$ tens of pc) dense clumps not resolved in cosmological simulations 
\citep[e.g.,][]{Cantalupo2014,Hennawi2015}; (\textit{ii}) The radial dependence of $\Gamma_{\rm HI}$ depends on the escape fraction of ionizing photons from both the central galaxy and surrounding galaxies. The escape fraction of ionizing photons, and its dependence on redshift and galaxy type, is still highly uncertain. Given these uncertainties, it is useful to have an analytic formalism which allows for an efficient exploration of the parameter space.
We describe both model ingredients in more detail in the following section.

\section{Characterising the CGM and $\Gamma_{\rm HI}(\lowercase{r})$}
\label{sec:model}

We divide this section into two parts: in the first, we describe our models for $\Gamma_{\rm HI}(r)$ 
[\S~\ref{sec:gamma}], and in the second, we describe our models for the CGM [\S~\ref{sec:cgm}].

\subsection{Modeling $\Gamma_{\rm HI}(r)$}
\label{sec:gamma}
We divide the total photoionization rate $\Gamma_{\rm HI}(r)$ at a distance $r$ from the central 
galaxy into ({\it i}) the contribution from the central galaxy itself, $\Gamma^{\rm cen}_{\rm HI}(r)$, 
and ({\it ii}) the contribution from all other ionizing sources, $\Gamma^{\rm ext}_{\rm HI}(r)$, i.e., 
\begin{equation}\label{eq:gammamaster}
\Gamma^{\rm tot}_{\rm{HI}}(r)=\Gamma^{\rm cen}_{\rm{HI}}(r)+\Gamma^{\rm ext}_{\rm{HI}}(r)~.
\end{equation} We discuss each component separately below.

\subsubsection{Modeling $\Gamma^{\rm cen}_{\rm HI}(r)$}\label{sec:gamma_cen}
The photoionization rate due to the central galaxy at distance $r$ is
\begin{equation}\label{eq:gamma}
\Gamma^{\rm cen}_{\rm{HI}}(r)=f_{\rm esc}(r)\times SFR \times \frac{{\rm \dot n_{\rm ion}}\,\sigma_{\rm{HI}}^{912}}{4\pi r^2}\frac{\gamma}{\gamma\,-\,3}~,
\end{equation}
where ${\rm \dot n_{ion}}=10^{53}\,{\rm s}^{-1}$ is the total number of ionizing photons emitted 
per unit SFR and time, where SFR denotes the star formation rate in ${\rm M_{\odot}}$ yr$^{-1}$.
$\sigma_{\rm{HI}}^{912}$ is the hydrogen photoionization cross-section at the Lyman limit frequency, 
$\nu_{912}$, which has a value $\sigma_{\rm{HI}}^{912}= 6.3\times10^{-18}\,{\rm cm^2}$. 
We assume that the spectral energy distribution (SED) of a galaxy near the Lyman limit is given by 
$L(\nu)=L_{912}\,(\frac{\nu}{\nu_{912}})^{\gamma}$,
where we take $\gamma=-2$ following the work of \cite{Becker2013}, although the value for this 
parameter is a matter of debate covering the range between -1 and -3 \citep[see  
discussions in][]{Kuhlen2012,Becker2013}. We include the radial dependence of 
the escape fraction of ionizing photons in terms of $f_{\rm c}(r)$ as
\begin{equation}\label{eq:fesc}
f_{esc}(r) = {\rm exp}\left[ -\int_0^r f_c(r)\,{\rm d}r\right] ~. 
\end{equation} 
For simplicity we assume that $f_{\rm esc}(r=0)=100\%$ and that the radial 
dependence of $f_{\rm esc}$ is determined entirely by the self-shielding clumps. We show in 
Figure \ref{fig:clumps} that in our models $f_{\rm esc}$ reduces to values that are in good 
agreement with existing observational constraints. We can account for additional absorption 
of ionizing photons in the ISM (or molecular clouds) of the galaxy by imposing 
$f_{\rm esc}(r=0)<$ 100\%. This restriction simply scales down our predictions by the same factor.

\subsubsection{Modeling $\Gamma^{\rm ext}_{\rm HI}(r)$}\label{sec:gamma_ext}
The photoionization rate due to the external galaxies surrounding the central 
LAE can be obtained by summing the contribution of all of them as 
\begin{equation}\label{eq:gammaext}
\Gamma^{\rm ext}_{\rm{HI}}(r)=\frac{\sigma_{\rm{HI}}^{912} \gamma}{4\pi(\gamma\,-\,3)}\sum_i~\frac{\dot{N}_{\rm ion,i}}{|\vec{{\bf r}} -\vec{{\bf r}}_i|^2}\exp{-\frac{|\vec{{\bf r}} -\vec{{\bf r}_i}|}{\lambda_{\rm mfp}}},
\end{equation} 
where the effective ionizing emissivity of galaxy `i' is defined 
as $\dot{N}_{\rm ion,i}\equiv  \dot{n}_{\rm ion,i}\times {\rm SFR}_{\rm i}\times f_{\rm esc,i}$,   
$\lambda_{\rm mfp}$ denotes the effective mean free path of ionizing photons, and 
where we assumed that each source of ionizing radiation has the same spectral slope $\gamma$. 
The continuous version of Eq.~\ref{eq:gammaext} can be written as 
\begin{align}\label{eq:gammaextcon}
\Gamma^{\rm ext}_{\rm{HI}}(r)=&\frac{\sigma_{\rm{HI}}^{912} \gamma}{4\pi(\gamma\,-\,3)}\int {\rm d}V^{\prime} 
\frac{\epsilon_{\rm LyC}(r^{\prime})}{|\vec{{\bf r}} -\vec{{\bf r}^{\prime}}|^2}\exp{-\frac{|\vec{{\bf r}} 
-\vec{{\bf r}^{\prime}}|}{\lambda_{\rm mfp}}} \nonumber \\
=&\frac{\sigma_{\rm{HI}}^{912} \gamma}{4\pi(\gamma\,-\,3)}\int_{0}^{\infty} r^{\prime 2}{\rm d}r^{\prime}\int_{0}^\pi \sin\theta d\theta  ~ ~ \nonumber \\
\times&\int_0^{2\pi} d\phi \frac{\epsilon_{\rm LyC}(r^{\prime})}{y(r^{\prime}, \theta, \phi)^2} 
\exp{-\frac{y(r^{\prime}, \theta, \phi)}{\lambda_{\rm mfp}}}  ~, ~ ~\nonumber \\ 
\end{align}
where $y=\sqrt{|r^{\prime 2}+r^2-2rr^{\prime}\sin\theta\sin\phi|}$. Here, $\epsilon_{\rm LyC}(r^{\prime})$ denotes the effective production rate of ionizing photons per unit volume. 
This quantity is poorly constrained as it depends on a number of 
factors including ({\it i}) the non-linear clustering of galaxies around the central galaxy as 
a function of their $L_{\rm UV}$, and ({\it ii}) the escape fraction of ionizing photons as a 
function of $L_{\rm UV}$\footnote{We can write out explicitly an expression for $\epsilon_
{\rm LyC}(r)$ as
\begin{equation}\label{eq:eps2}
\epsilon_{\rm LyC}(r)=\int_0^\infty {\rm d}L_{\rm UV}\frac{{\rm d}n}{{\rm d}L_{\rm UV}}[1+\xi(r,L_{\rm UV})]\dot{n}_{\rm ion}(L_{\rm UV})f_{\rm esc}(L_{\rm UV}) \nonumber
\end{equation} 
where $\frac{{\rm d}n}{{\rm d}L_{\rm UV}}$ denotes the UV-luminosity function (number 
density of galaxies in the range $L_{\rm UV} \pm {\rm d}L_{\rm UV}$), $\dot{n}_{\rm ion}(L_{\rm UV})$ 
denotes the production rate of ionizing photons for a galaxy with $L_{\rm UV}$, and $f_{\rm esc}$ 
denotes the escape fraction of ionizing photons.}. We can express $\epsilon_{\rm LyC}(r)$ as 
\begin{equation}\label{eq:eps2}
\epsilon_{\rm LyC}(r)=\langle \epsilon_{\rm LyC}\rangle [1+\xi_{\rm LyC}(r)],
\end{equation} 
where $\langle \epsilon_{\rm LyC}\rangle$ denotes the cosmic average 
production rate of ionizing photons per unit volume. Importantly, $\langle \epsilon_{\rm LyC}\rangle$ 
can be obtained 
directly from observations, by combining measurements of $\lambda_{\rm mfp}$ and 
photoionization rates from the Ly$\alpha$ forest \citep[see, e.g.,][]{Kuhlen2012}. The use of 
$\langle \epsilon_{\rm LyC}\rangle$ allows us to circumvent the above-mentioned problems 
related to the integration limits for the luminosity function and the use of assumptions for the 
escape fraction of ionizing photons. The quantity $\xi_{\rm LyC}(r)$ is the two point 
correlation function and quantifies how ionizing sources are clustered around the central galaxy, 
and is clearly important in our analysis. In the next section, we present 
how we calculate $\xi_{\rm LyC}(r)$, and discuss other model assumptions.

\subsubsection{Model Assumptions}\label{sec:assump}
\begin{table}
	\begin{center}
	\caption{Parameters for the calculation 
	of the ionizing background.\label{ta:params}}	
	\begin{threeparttable}
		\begin{tabular}{cccc} 
		\tableline \\
		z		   &$\langle\epsilon_{\rm LyC}\rangle\tnote{a}$        &$f_{esc}\tnote{b}$      		&$\lambda_{\rm mfp}\tnote{a}$ 	      \\ 
				   &${\rm [10^{50}\,s^{-1}\,cMpc{-3}]}$					&								& [pMpc]		\\  \tableline  \\
		3.1	      	 &$2.7$     						 		&$0.026$                     				&$84.4$       		        \\ 
		\tableline	
		\end{tabular}
		\begin{tablenotes}
			\item[a] Values from Table 2 in \cite{Kuhlen2012}. 
			\item[b] Average escape fraction of ionizing photons computed using Eq. (14) in \cite{Kuhlen2012}, with the parameters
				$\zeta_{ion}=1$,  $\kappa=2$ and assuming an $f_{esc}(z=4) = 0.04$.
		\end{tablenotes}
	\end{threeparttable}
	\end{center}
\end{table}
Our approach is the following: 
\begin{itemize}[leftmargin=0pt,itemindent=20pt]
\item For the central LAE we assume a SFR=$10\hspace{1mm}{\rm M_{\odot}}$ yr$^{-1}$.
\item We use for $\lambda_{\rm mfp}$ and $\langle \epsilon_{\rm LyC}\rangle $, the values 
shown in Table \ref{ta:params}. These values are taken from the work of \cite{Kuhlen2012} 
who compilated them from \cite{FaucherGiguere2008,Songaila2010,Prochaska2009,Bolton2007}. 
\item For the integral over $r^{\prime}$ in Eq.~\ref{eq:gammaextcon},
we set the lower limit $r^{\prime}_{min} = 10\, {\rm pkpc}$.  We 
tested the value $r^{\prime}_{min} = 5\, 
{\rm pkpc}$ and we have seen that it does not change our results significantly. The upper 
limit is set to $r^{\prime}_{max} = 
100\,\lambda_{\rm mfp}$, considering the effect of very distant galaxies. We have 
explicitly tested  that the effect of considering a larger limit does not change the results. 
\item We consider that the real space two-point correlation function for ionizing 
sources, $\xi_{\rm LyC}(r)$, can be described as 
\begin{equation}\label{eq:xi}
\xi_{\rm LyC}(r)=b_{\rm LyC}\,b_{\rm LAE}\,\left[ \xi(r)_{\rm 1h}+\xi(r)_{\rm 2h}\right]~ ,
\end{equation} 
where we have used that the cross correlation function of two different tracers, each with 
its own bias, is given by $\xi_{\rm ab}(r)=b_{\rm A}\,b_{\rm B}\,\xi(r)$. In our case, the first tracer 
is an LAE, and we take its scale depending bias, $b_{\rm LAE}(r)$, from fitting the data in the
work of \cite{Ouchi2010} (See Appendix \ref{sec:bias} for a description of the procedure). The 
bias $b_{\rm LyC}$ denotes an `effective' bias of LyC emitting sources. Since we currently
do not know which galaxies are LyC sources, we leave this quantity as a free parameter, 
proportional to the bias of LAEs by a constant. Our fiducial calculation assumes 
$b_{\rm LyC}(r) = b_{\rm LAE}(r)$ but we also discuss the effect of considering other values.
Finally, the terms $\xi_{\rm 1h}(r)$ and $\xi_{\rm 2h}(r)$ denote the 1-halo and 2-halo terms of 
the two-point correlation function of dark matter. The 2-halo term accounts for clustering of 
matter in different dark matter halos, while the 1-halo term accounts for clustering of matter 
within the same halo. In the Appendix \ref{sec:corr}, we detail the calculations for the obtention 
of these correlation functions.
 \item \cite{Matsuda2012} showed that the LAE overdensity is 
correlated with the spatial extent of Ly$\alpha$ halos around LAEs. We compute 
the overdensity of LAEs, $\delta_{\rm LAE}$, within a sphere of $R=2$ Mpc h$^{-1}$ around our central galaxy as
\begin{equation}
~ \delta_{\rm LAE}=\frac{n_{\rm LAE}-\bar{n}_{\rm LAE}}{\bar{n}_{\rm LAE}}=\frac{3}{R^3} 
\int_0^R r^2{\rm d}r\hspace{1mm} \xi_{\rm LAE}(r) ~ , 
\end{equation} 
where $n_{\rm LAE}$ is the number of LAEs in the overdense region and $\bar{n}_{\rm LAE}$ 
is the average value in the field. We obtain a value for the overdensity  
$\delta_{\rm LAE} \sim 1.5$, which will be important when comparing our results with other 
works.
\item For simplicity, we assume that the escape fraction of Ly$\alpha$ 
photons is $f_{\rm esc}^{Ly\alpha}=100\%$. For the case of fluorescence, Ly$\alpha$ 
photons escape from a medium with a column density N$_{\rm HI}\sim 10^{17}$ cm$^2$. For 
such low column densities the effects of dust are expected to be small \citep[e.g.,][and 
references therein]{Gronke2015}. 
Our results can be scaled directly with $f_{\rm esc}^{Ly\alpha}$ assuming that the escape 
fraction is not spatially dependent.

\end{itemize}

\subsection{Modeling the CGM}\label{sec:cgm}
Below we describe how we parametrize the CGM. We construct models 
of the CGM 
using simplified clumpy outflow models following \cite{Steidel2010} in \S~\ref{sec:clumps}, 
and \cite{DijkstraKramer2012} in \S~\ref{sec:dk}. In \S~\ref{sec:rahmati} we take an 
alternative approach, and adopt a fitting formula for the `area' covering factor of self-shielding 
gas in the cosmological hydrodynamical EAGLE simulations \citep{Rahmati2015}.

\subsubsection{Steidel et al. (2010) - Clumpy Outflow Model}
\label{sec:clumps}

We follow \cite{Steidel2010} for this model, where optically thick cold clumps 
are embedded within an outflowing optically thin hot medium, lying within the radial 
range $r_{min}=1\, {\rm pkpc}$ and $r_{max}={\rm 250\, pkpc}$. The clumps have 
a constant mass and are in pressure equilibrium with the hot medium, the pressure of the 
latter varying with distance as 
$p(r)\,\propto\,r^{-2}$ \citep{Steidel2010}. 

The covering factor, $f_{\rm c}(r)$, is given by \citep{DijkstraKramer2012} 
\begin{equation}\label{eq:fcovclump}
f_{\rm c}(r)=n_{\rm c}(r)\,\sigma_{\rm c}(r),
\end{equation} where $n_{\rm c}(r)$ denotes the number density of clumps at $r$ and 
$\sigma_{\rm c}(r)$ denotes their geometric cross-section. These quantities are given by 
\begin{eqnarray}\label{eq:ncsc}
\sigma_c(r)=&\pi\,R_c(r)^2 ~,\\ \nonumber
n_c(r)=&\frac{C_n}{4\,\pi\,r^2\,v_c(r)} ~,
\end{eqnarray}
where $R_{\rm c}(r)$ denotes 
the cloud radius. 
Pressure equilibrium dictates that $R_c = C_r\, r^{-2/3}$. The constant of 
proportionality $C_r$ is constrained by the observations, and depends on the number of clumps 
\citep[see][for more details]{DijkstraKramer2012}. We adopt the value $C_r = 0.01$. 
$C_n$ is again a free parameter which we set to $\rm C_n=10^{-10}\,{\rm s^{-1}}$, in order 
to obtain agreement with the observations in \cite{Steidel2010}. For the outflow velocity profile, 
we assume that the clumps undergo a radial acceleration of the form 
$a_c(r)=Ar^{-\alpha_v}$. Under this assumption, the radial velocity profile is \citep{Steidel2010}
\begin{equation}\label{eq:vel}
v_c(r)=\left(\frac{2A}{\alpha_v\,-\,1}\right)^{1/2}\left(r_{min}^{1-\alpha_v}\,-\,r^{1-\alpha_v}\right)^{1/2}~,
\end{equation}
for ${\alpha_v} > 1$. We do not take into account values ${\alpha_v} \le 1$ in this work. 
Furthermore, $A$ is a constant that sets the velocity at large radius, $r \rightarrow \infty$, 
through $v_\infty=\sqrt[]{2\,A\,r_{min}^{1-\alpha_v}/(\alpha_v\,-\,1)}$. Following \cite{Steidel2010} 
we set $v_\infty=\,830\,{\rm km\,s^{-1}}$ and $\alpha_v =\,1.4$.  See Figure 2 in 
\cite{DijkstraKramer2012} and Figure \ref{fig:clumps} in the Appendix \ref{sec:clumpfig} for the 
visualization of these parameters.

We note that this simple model was designed to reproduce CGM absorption line data  
from composite Lyman Break Galaxy (LBG) spectra. 
For simplicity, we will use this model for the environment of LAEs. This may artificially enhance the 
cold gas content of our CGM, and we will return to this when we discuss our results. Also, this model 
assumes spherical symmetry, which does not represent the true complexity of CGM 
\citep[e.g.,][]{Gauthier2012}.

\subsubsection{Dijkstra \& Kramer (2012) - Decelerated Outflow}
\label{sec:dk}

In the model presented by \cite{DijkstraKramer2012} the outflow velocity 
decreases beyond a certain radius. This modification allowed \cite{DijkstraKramer2012} to 
simultaneously reproduce absorption and emission around LBGs with a simple scattering model. 
Their velocity profile was constructed by assuming that the radial clumps undergo an additional 
radial deceleration due to gravity, which dominates beyond some 
`transition' radius. The resulting velocity can be written as
\begin{equation}
v_c(r)=2\sigma \sqrt{\ln\left(\frac{r_{\rm min}}{r}\right)+\frac{A}{2\sigma^2(1-\alpha_v)}\left(r^{1-\alpha_v}-r_{\rm min}^{1-\alpha_v}\right)}    ~,
\end{equation}
where $\sigma$ denotes the velocity dispersion and the other parameters 
denote the same as in the Steidel et al. model. We adopt for our model parameters the 
same values as in model IV in \cite{DijkstraKramer2012}, to whom we refer 
the reader for a more detailed discussion about the model.
The reduced outflow velocity translates to a larger clump number density (Eq.~\ref{eq:ncsc}), and thus enhanced covering factor, which in turn affects the radial dependence 
of $f_{\rm esc}(r)$ (Eq.~\ref{eq:fesc}).


\subsubsection{Rahmati et al. (2015) - Fit to EAGLE Simulations}
\label{sec:rahmati}

 \cite{Rahmati2015} provided a fitting formula for the {\it area covering fraction} of 
self-shielding gas around simulated galaxies in the EAGLE simulation. This `area covering 
factor', $F_{\rm LLS}$, denotes the fraction of the total area $2\pi b{\rm d}b$ at impact parameter 
$b$ that is covered by this gas. We can interpret this fraction $F_{\rm LLS}$ as the probability 
that a sightline at impact $b$ intersects a self-shielding cloud, or more accurately, at least 
one self-shielding cloud. In other words, we can set $F_{\rm LLS}(b)=P(N_{\rm clump} 
\geq 1, b)$ (which is the quantity we introduced in \S~\ref{sec:formalism}). The fitting formula 
in \cite{Rahmati2015} is
\begin{equation}
F_{\rm LLS}(x)=1-\frac{1}{1+\left(\frac{L_z}{x} \right)^{\alpha}}+C\left[ \frac{1}{1+\left( \frac{L_z}{x}\right)^3}\right]10^{\frac{z-4}{3}},
\end{equation} 
where $x\equiv r/r_{\rm vir}$, $z$ denotes redshift and $L_z=AB^z$, in which $A,B,C$ and $\alpha$ 
are parameters that depend on the type of absorber that is considered.
We use the parameters presented in Table 2 of \cite{Rahmati2015} for the case of Lyman Limit 
Systems (LLS) and velocity width $\Delta{\rm v} = 3000\, {\rm km/s}$. 
We note that the 
simulations by \cite{Rahmati2015} only include the global ionizing background by 
\cite{Haardt2001}, i.e., this 
fitting formula ignores that the photoionization rate in close proximity to galaxies is elevated. 
The authors argue, however, that this assumption only affects the fitting formula at the $~10\%$
level at distances below the virial radius, and we therefore adopt it as it is. 

We now use that $ N_{\rm clump}(b) $ and $f_{\rm c}(r)$ are related to 
each other via an Abel-transform (see Eq.~\ref{eq:nclump}). We can invert this transformation 
and obtain $f_{c}(r)$ from $N_{\rm clump}(b) $ as
\begin{equation}\label{eq:abel}
f_{\rm c}(r)=-\frac{1}{\pi}\int_{r}^{\infty}\frac{{\rm d} N_{\rm clump} }{{\rm d}y}\frac{{\rm d}y}{\sqrt{y^2-r^2}} ~,
\end{equation} 
where we obtain $\langle N_{\rm clump} \rangle$ from $F_{\rm LSS}=P(N_{\rm clump} 
\geq 1, b)$ by inverting Eq.~\ref{eq:Nclos} to get
\begin{equation}
N_{\rm clump}(b) =-\ln{\left(1-F_{\rm LLS}\right)}~.
\end{equation} 
We can thus convert the fitting formula from \cite{Rahmati2015} into $f_{\rm c}(r)$, 
once we specify the virial radius $r_{\rm vir}$. For our model we adopt  
$r_{\rm vir}=80\, {\rm pkpc}$  accounting for the 
values found for host dark matter halos of ${\rm M_h\sim10^{12}\, M_{\odot}}$ 
\citep[e.g.,][]{Cooke2013}, which corresponds to the lowest mass for which the fitting 
formula is valid. Finally, the upper integration limit in Eq.~\ref{eq:abel} is $r 
\rightarrow \infty$. However, in practice we use $r_{max}={\rm 250\, pkpc}$, which 
corresponds to the maximum distance at which cold clumps exist in our model,  
following \cite{Steidel2010}. We have verified that this consideration does not affect 
our main results. Figure \ref{fig:clumps} in the Appendix shows the values of the 
escape fraction and the covering fraction for the case of this method.


\section{Ionization and surface brightness profiles}\label{sec:results}

We present the results for $\Gamma(r)$ in \S~\ref{sec:gammar} 
and $SB_{Ly\alpha}$ in \S~\ref{sec:sbprofile}. We restrict our results to 
distances larger than 10 pkpc since we consider that below this value, the
physics governing the processes within the central galaxy is not accurately represented 
by our simple methodology.

\subsection{Radial Profile $\Gamma_{\rm HI}(r)$}
\label{sec:gammar}

\begin{figure*}
\begin{center}
\includegraphics[width=0.34\textwidth]{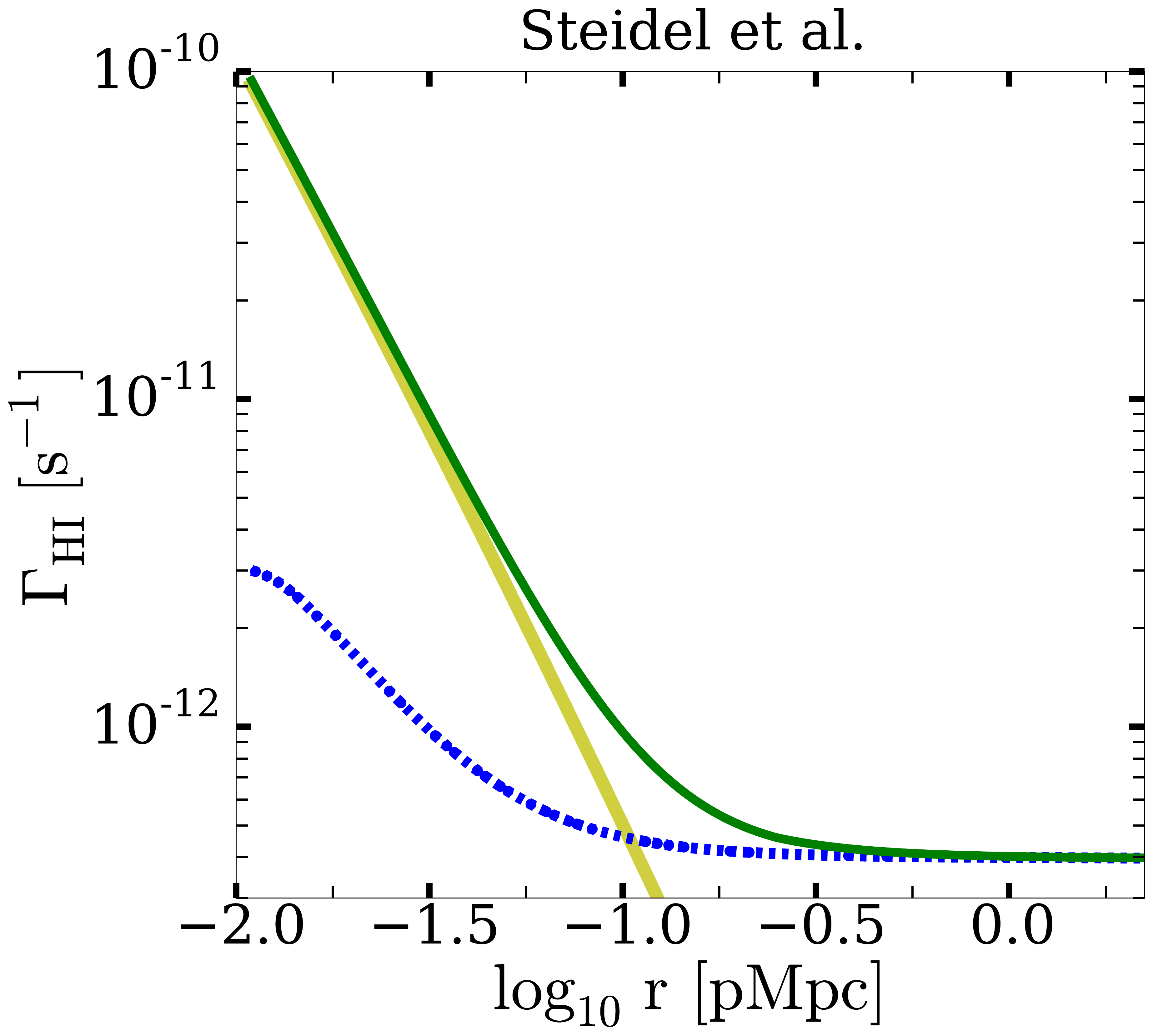}\includegraphics[width=0.32\textwidth]{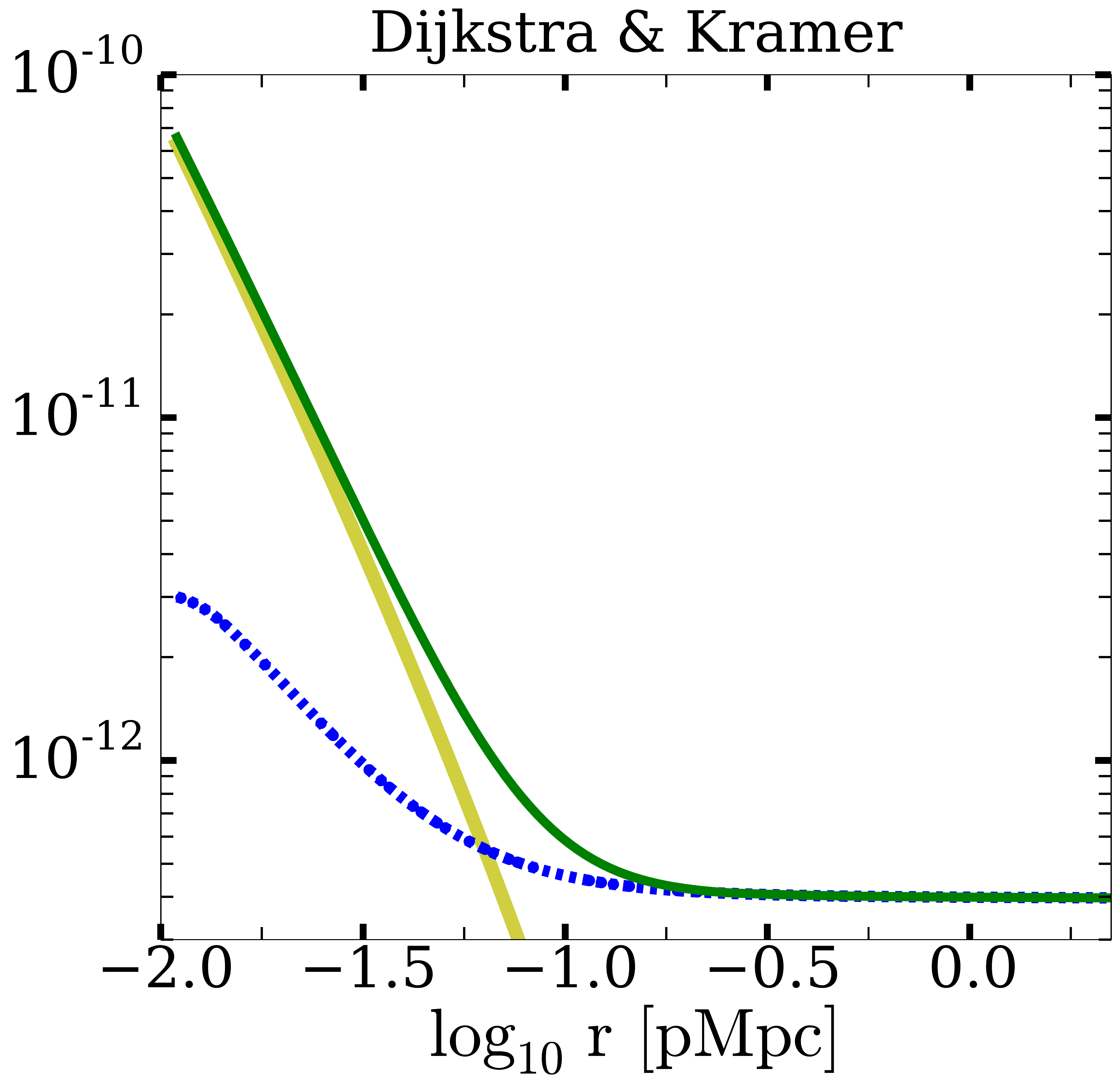}\includegraphics[width=0.32\textwidth]{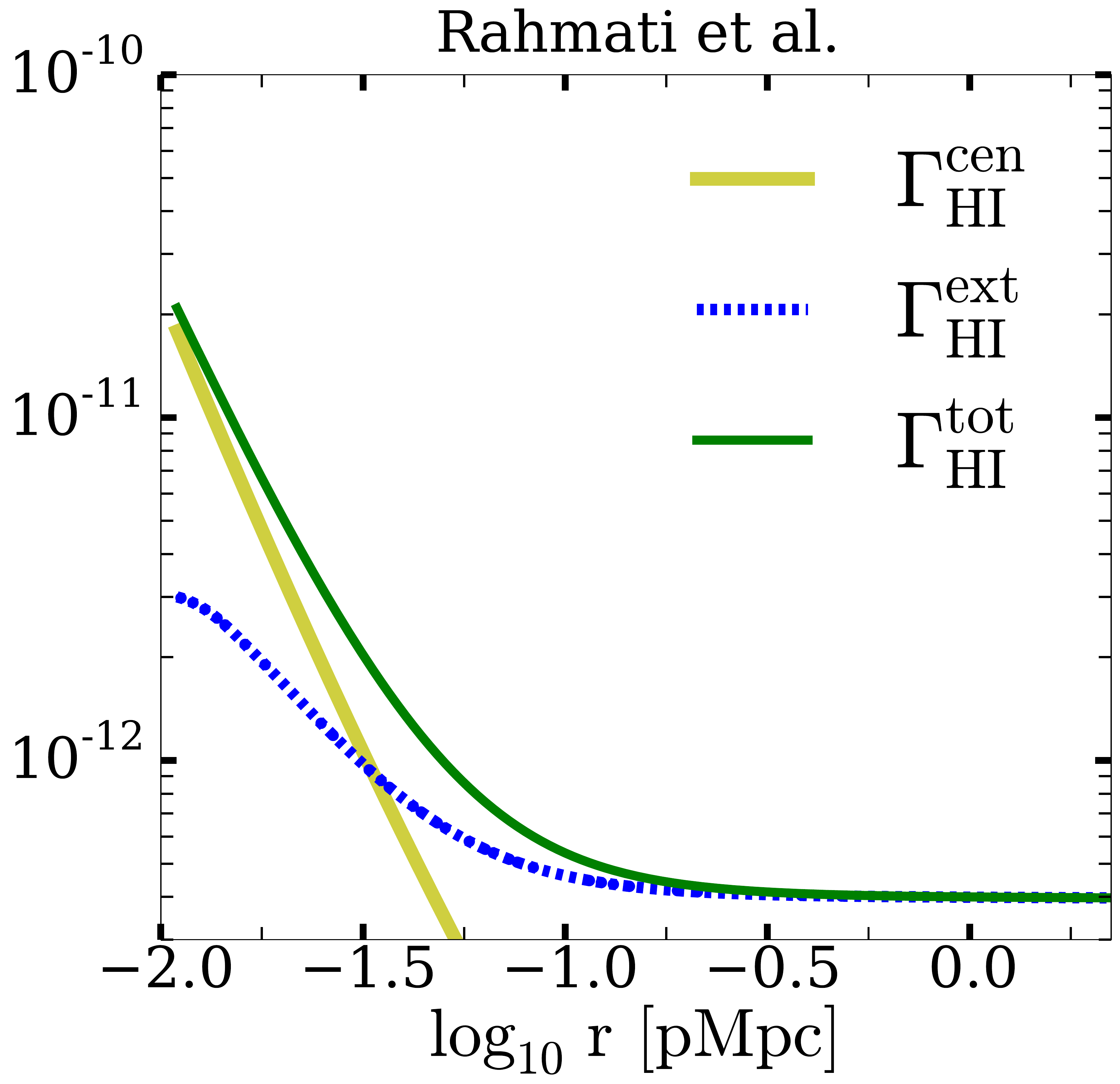}
\caption{Enhancement of the ionizing radiation with radial distance around the central LAE, expressed 
in terms of the photo-ionization rate, using $b_{\rm LyC}(r)=b_{\rm LAE}(r)$ for our three models. 
The {\it thick yellow line} indicates the contribution of the central LAE. The {\it dashed 
blue line} denotes the effect produced by the external galaxies and the {\it continuos green line} 
shows the addition of the two. Note the small contribution of the external galaxies to the boost, 
except for the case of the Rahmati et al. model. At a distance $r\sim 10\, {\rm pkpc}$, the boost 
in the ionizing background reaches a factor of $\sim$50 for the Rahmati et al. case and $\gsim 
200$ for the other two models.} 
\label{fig:ionbg}
\end{center}
\end{figure*}

We show in Figure \ref{fig:ionbg} the enhancement of the ionizing background in the 
medium surrounding the central LAE for our three models. The {\it thick yellow solid line} 
shows the contribution from the central galaxy and 
the {\it blue dotted line} denotes the contribution from the external galaxies, where we 
adopted $b_{\rm LyC}(r)=b_{\rm LAE}(r)$ and considered non-linear clustering. The 
{\it green solid line} shows the total photoionization rate, which is simply the sum of the 
central and external contributions. 

Figure \ref{fig:ionbg} shows that photoionization by the central source 
dominates at small distances for all models. At precisely what radius external galaxies contribute 
equally depends on the model. Most notably, external galaxies start dominating closer to the 
central galaxy (at $r\sim30\,{\rm pkpc}$) in the Rahmati et al. model because the 
photoionization rate from the central galaxy lies significantly lower than in the other two 
models. This is because in this model $f_c(r)$ is larger at small $r$, which causes 
$f_{\rm esc}$ to have dropped to a lower value at $r=10$ pkpc compared with the other two models.

In any case, in all models the enhancement of the photoionization rate is substantial:  
a factor $\gsim 10$ at $r\lsim 30\, {\rm kpc}$, and goes up to a factor of $\sim50$ ($\gsim 250$) 
at $r\sim 10\, {\rm pkpc}$ for the Rahmati et al. model (Steidel et al. and Dijkstra \& Kramer 
models). At larger $r$, we recover the average value for the ionizing background, 
although we find the value $\Gamma_{\rm HI}\sim4\times10^{-13}\, {\rm s^{-1}}$, 
which is slightly smaller than the result obtained by \cite{Kuhlen2012}, $\Gamma_{\rm HI}\sim6
\times10^{-13}\, {\rm s^{-1}}$.  This difference arises from the value for the spectral slope,  
which we adopted from \cite{Becker2013}  for our calculations (see \S~\ref{sec:gamma_cen}),  
but it does not affect our results significantly.

\subsection{Radial Profile Surface Brightness $SB_{\rm Ly\alpha}(b)$}
\label{sec:sbprofile}
\begin{figure*}
\begin{center}
\includegraphics[width=0.34 \textwidth]{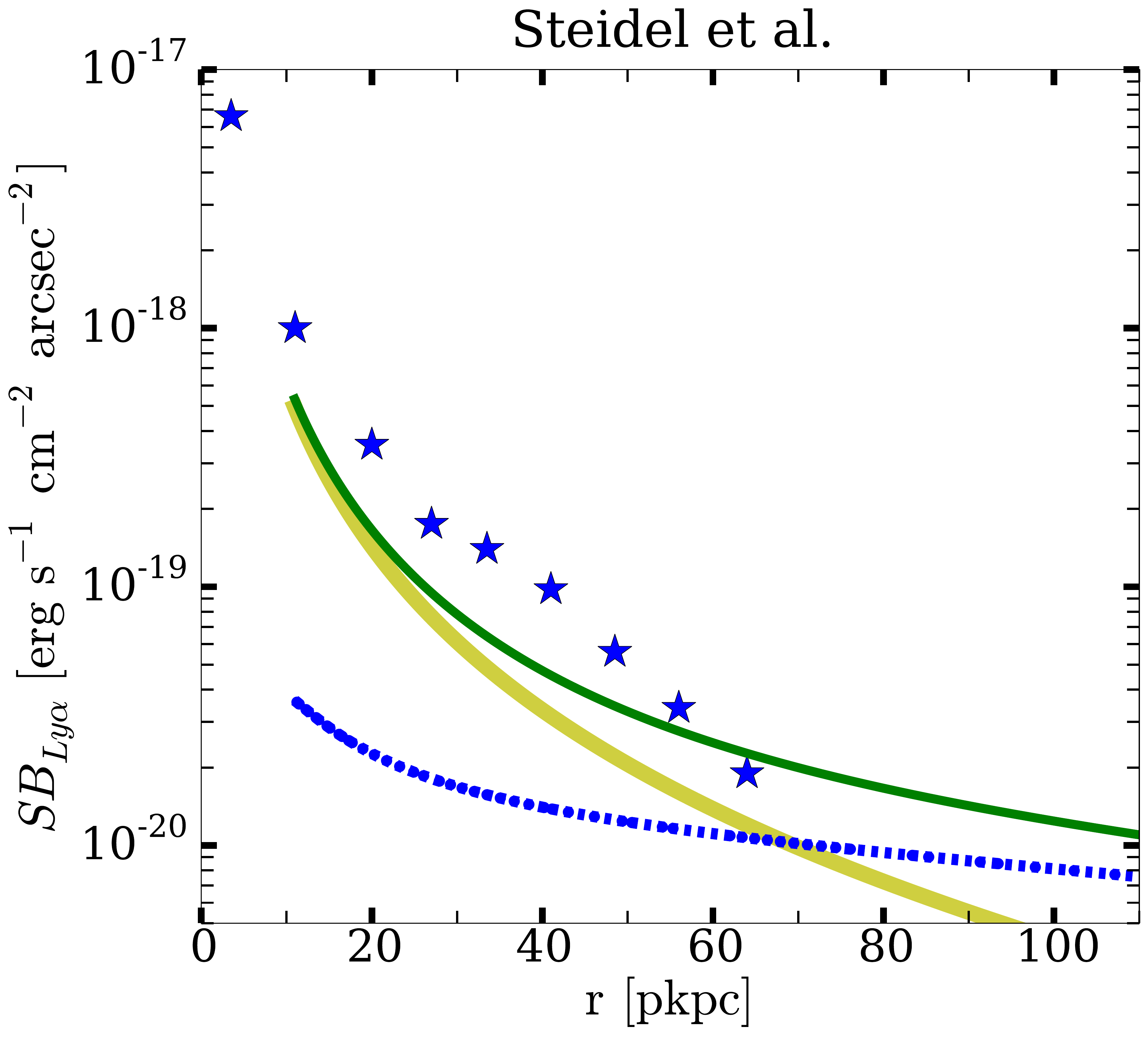}\includegraphics[width=0.32 \textwidth]{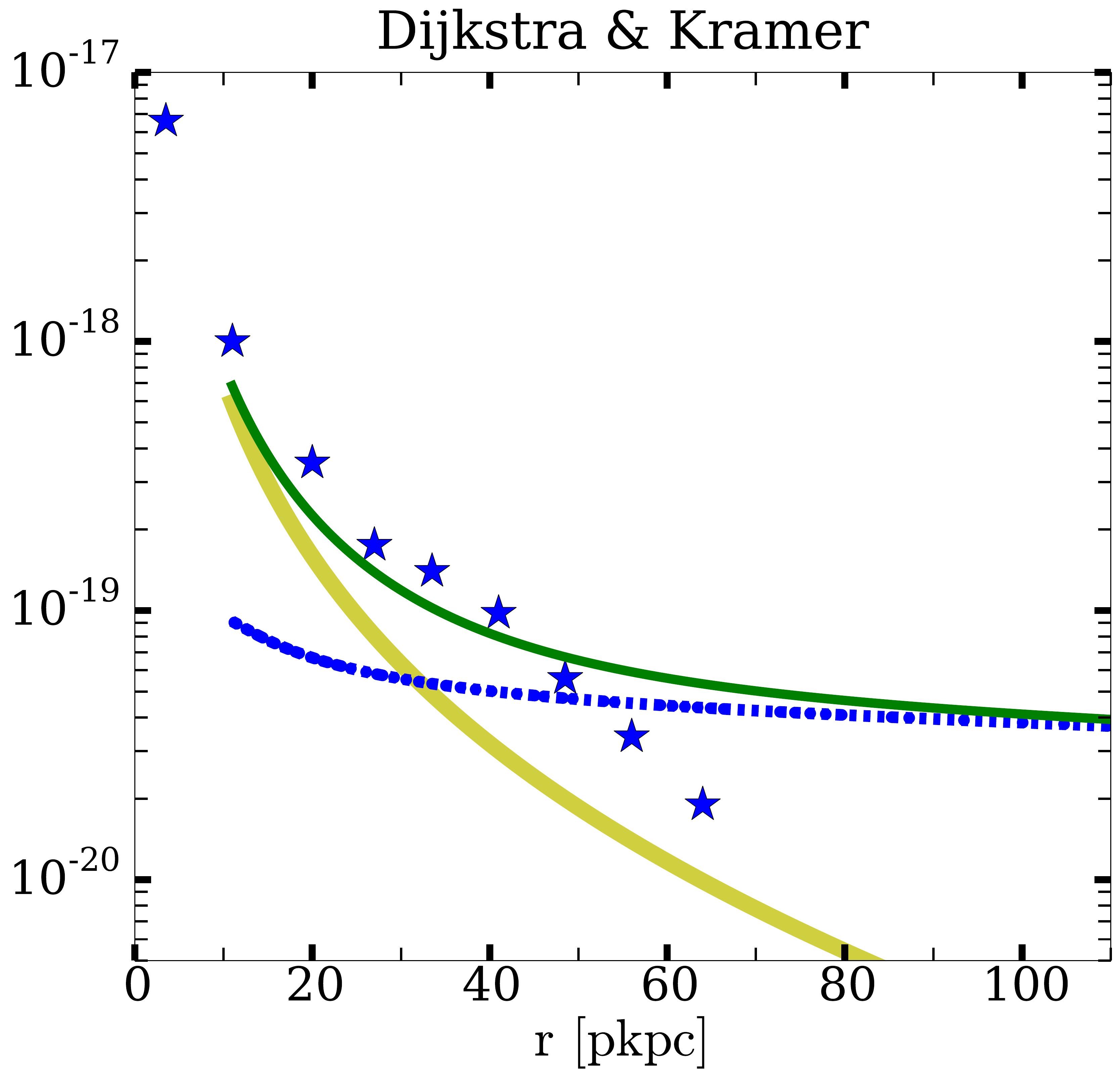}\includegraphics[width=0.32 \textwidth]{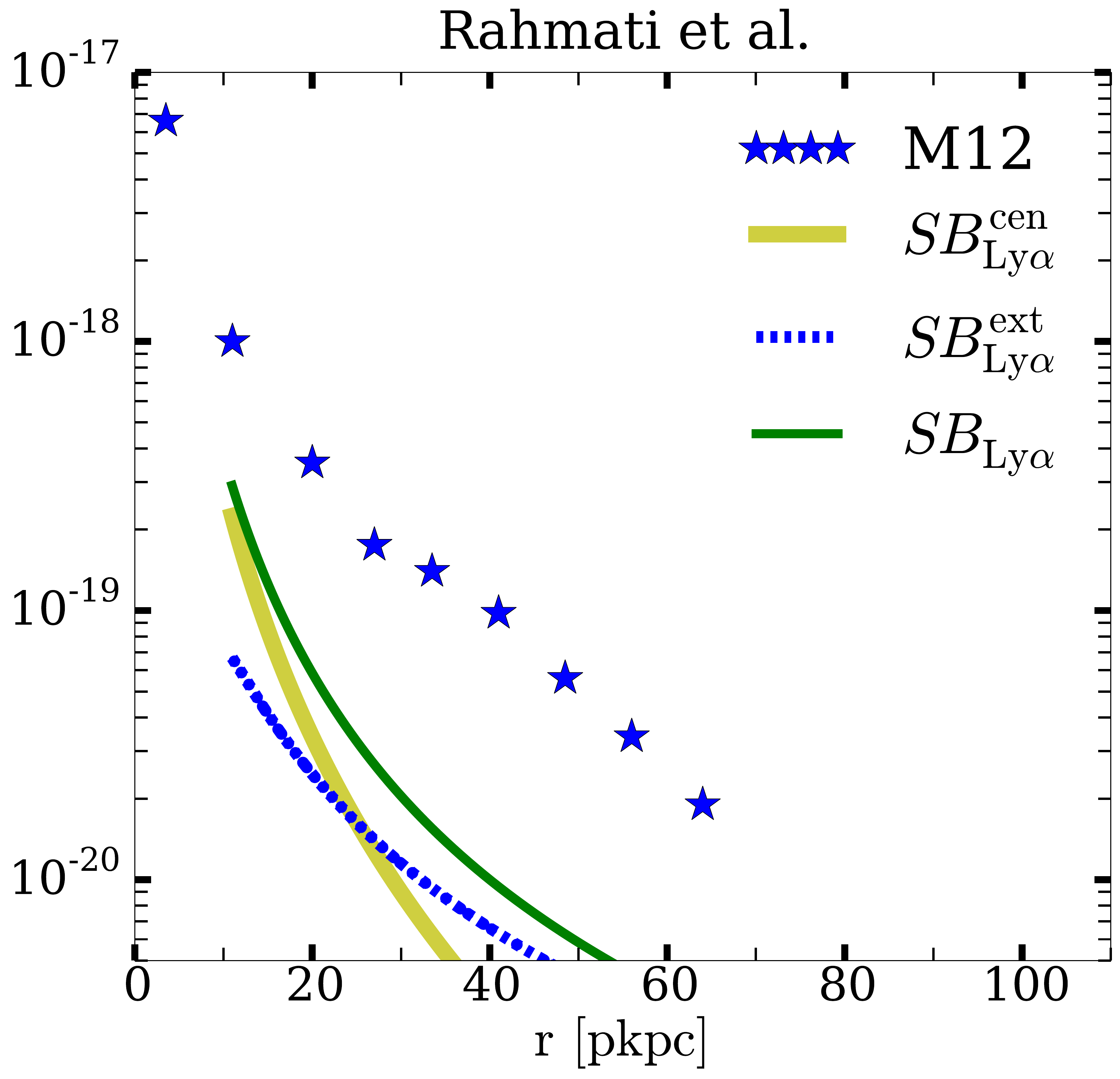}
\caption{Ly$\alpha$ surface brightness profiles for our three CGM models assuming 
$b_{\rm LyC}(r)=b_{\rm LAE}(r)$ and $f_{\rm esc}^{Ly\alpha}=100\%$. The x-axis denotes the 
impact parameter in physical units. The {\it thick yellow 
line} shows the profile due to the central LAE and the {\it blue dashed line} the contribution from 
the external galaxies. The {\it continuos green line} shows the total surface brightness profile and 
the stars denote a simple estimation of the data in \cite{Matsuda2012}, for their overdensity range 
$2.5<\delta_{LAE}<5.5$. The Steidel et al. and Dijkstra \& Kramer models, where the effect of the 
central galaxy is very important, are close to the observed data. The latter, however, shows a signal 
at large distances, due to the effect of the external galaxies, above the observations. The Rahmati 
et al. model is well below the data and only the central galaxy approaches the observations at very 
small distances.}
\label{fig:sbz3}
\end{center}
\end{figure*}

Figure~\ref{fig:sbz3} shows the Ly$\alpha$ surface brightness profiles for our 
three models, as given by Eq.~\ref{eq:sb}. For comparison, we show a simplified view of 
the data for the highest density range, $2.5<\delta_{LAE}<5.5$, in \cite{Matsuda2012} 
(their upper-left panel in Figure 3).

The {\it left panel} shows the Steidel et al. model, which predicts a surface brightness 
that lies a factor $\sim1.5$ below the observations at $10<r<30$ 
pkpc. This factor 
varies slightly at larger distances but the general behaviour 
is well reproduced. For this model, the central galaxy dominates the profile up to $r\sim70$ pkpc 
and the contribution from external galaxies is only important at larger distances. However, the strong impact of 
the central galaxy requires a very high ionizing photon escape fraction at distances below 50 pkpc
($\sim 30\%$; Figure \ref{fig:clumps}).
The {\it middle panel} of Figure \ref{fig:sbz3} shows the Dijkstra \& Kramer model. 
The high covering factor, $f_c$, at large distances yields a profile which is significantly above 
the observed data (a factor $\sim3$ at $r\sim65$ pkpc).
The {\it right panel} denotes the Rahmati et al. model which presents a profile significantly below 
the observed data, a factor $\sim$3 ($\sim$8) at $r\sim10$ ($r\sim65$) pkpc. This is 
because in this model $f_c$ and $f_{\rm esc}$ are much smaller at large $r$ (see Figure 
\ref{fig:clumps}), which reduces the surface brightness. 
We note that the Steidel et al. and Dijkstra \& Kramer models were designed to 
reproduce observations of (the more massive) LBGs, and may therefore overpredict the 
surface brightness.

\cite{Rahmati2015} formally adopted an ionizing background that was $\sim3$ times higher 
than ours when calculating $F_{\rm LLS}{\rm(b)}$, which affects their covering fraction at large 
distances. We note however, that the small differences that this introduces in the predicted 
fluorescent surface brightness are sub-dominant to differences introduced by local enhancements 
in the ionizing background.

We remind here that our overdensity value is $\delta_{LAE}\sim1.5$ and we are 
comparing to the range $2.5<\delta_{LAE}<5.5$ in \cite{Matsuda2012}. In \S~\ref{sec:rah} 
in the Appendix  
we show the results obtained when simply changing the value for the LAE bias (setting it to 
twice its default value) to obtain an overdensity $\delta_{LAE}\sim 5.8$. Also, we show 
the case when setting $b_{\rm LyC}(r)=3\,b_{\rm LAE}(r)$ to test the effects of a different bias 
for the ionizing sources. However, as we show, these variations do not affect our main 
results.


\section{Discussion of the Model}\label{sec:discussion}
There are several effects that we did not include in our modeling. First, our analysis assumes that the self-shielding clumps and ionizing sources are spatially uncorrelated. This assumption is likely unrealistic: for realistic cosmological density fields there are azimuthal density fluctuations at fixed galacto-centric radius $r$, and we expect both ionizing sources and self-shielding gas to preferentially reside in higher density regions\footnote{An example of such an azimuthal density fluctuation occurs when both the self-shielding gas and the ionizing sources reside in filaments that connect to the central galaxy}. If ionizing sources and self-shielding gas are spatially correlated, then this boosts the fluorescent signal: suppose that the local ionizing flux that is ``seen'' by a self-shielding cloud is due to a single nearby source, and that the spatial correlation between ionizing sources and self-shielding gas reduces the separation $d$ between the nearest ionizing source and the self-shielding gas by a factor of $x$ (i.e. $d \rightarrow xd$, where $x< 1$), then this boosts the local ionizing flux by a factor of $x^{-2}$. This can be a big effect. We note however, that by introducing a spatial correlation between the self-shielding gas and ionizing sources, we force the observed Ly$\alpha$ emission to closely trace the sources of ionizing radiation, which likely causes this model to predict that the non-ionizing UV continuum surface brightness profile should closely trace that of Ly$\alpha$. This is in disagreement with 
the observations in overdense regions by \cite{Steidel2010, Matsuda2012} although  
\cite{Momose2015} argues that a possible UV signal from the external galaxies may
be within the current detection threshold.

Second, we assumed that Ly$\alpha$ photons are observed from the location where they were produced. In reality we may expect Ly$\alpha$ photons to also scatter off the self-shielding gas clouds. Scattering redistributes the emission in the halo, and serves to flatten the predicted surface brightness profile \citep[e.g.,][]{Zheng2011,Hansen2006,DijkstraKramer2012,Laursen2007}.

Third, we have (purposefully) ignored other in-situ Ly$\alpha$ production mechanisms. 
Clustering of 
ionizing sources surrounding the central galaxy may also give rise to clustering of Ly$\alpha$ 
emitting sources around the central galaxy, as the sources of ionizing radiation are also sources 
of Ly$\alpha$ photons. We will address 
the contribution of satellite galaxies to the extended Ly$\alpha$ emission in a follow up paper.
Our main formalism can be easily modified to quantify this contribution.   

Fourth, we have ignored the contribution of possible cooling radiation (Ly$\alpha$ emission produced via collisional excitation in the neutral gas in response to, e.g., gravitational heating). We note that LAEs are thought to reside in low mass halos \citep[$M_h\sim10^{11.5}\,{\rm M_{\odot}}$, e.g.,][]{Gawiser2007}. Predicted Ly$\alpha$ cooling luminosities is a difficult problem which carries several uncertainties 
\citep[e.g.,][]{DijkstraLoeb2009, FaucherGiguere2010, Cantalupo2012}. Predictions for $M_h\sim10^{11.5}\,{\rm M_{\odot}}$  are that $L_{Ly\alpha} \sim 10^{42}$ erg s$^{-1}$  
\citep[][]{DijkstraLoeb2009, FaucherGiguere2010, Goerdt2010,Rosdahl2012}, and can reach 
a surface brightness $\gsim 10^{-19}$ erg s$^{-1}$ cm$^{-2}$ arcsec$^{-2}$ at $r\approx 20$ pkpc \citep[see Fig~13 of][]{Rosdahl2012} .

\cite{Lake2015} reproduced the Ly$\alpha $ surface brightness profile around 
LAEs naturally in their cosmological hydrodynamical simulations through a combination of 
cooling and satellite galaxies. Given the above mentioned uncertainties related to CGM 
modeling, it would be interesting to explore the constraints that this places on various subgrid prescriptions in the simulations. We will investigate the contribution from satellites to the Ly$\alpha$ surface brightness in a parameterized way in the companion paper. This will complement the work by Lake et al. (2015).


\section{Summary and conclusions}\label{sec:conclusion}

Fluorescence refers to the conversion of ionizing photons into Ly$\alpha$ on the surface of 
self-shielding gas in the circum-galactic medium (see \S~\ref{sec:intro}). 
We have computed whether fluorescence can explain spatially extended Ly$\alpha$ halos 
which have been ubiquitously observed around star forming galaxies. 

We have presented a general formalism for calculating the signal from fluorescence in 
\S~\ref{sec:formalism}. This formalism shows that there are two key model ingredients: 
({\it i}) the distribution of self-shielding gas in the circum-galactic medium, which is quantified 
via its (differential) covering factor, $f_c(r)$, which denotes the number of self-shielding clumps 
per unit length; ({\it ii}) the local ionizing radiation field, which is quantified by the photoionization 
rate $\Gamma(r)$. Here, $r$ denotes the galacto-centric distance from the galaxy. 
We argued in \S~\ref{sec:formalism} that it is not possible yet to model/simulate these 
quantities from first principles. It is therefore useful to have an analytic formalism which allows 
for an efficient exploration of the parameter space.

We presented our model for $\Gamma(r)$ in \S~\ref{sec:gamma}, which 
accounts for the distance dependent escape fraction of ionizing flux from the central 
galaxy. We also consider 
clustering of ionizing sources around the central galaxy, by boosting the {\it observationally inferred} volumetric production rate of ionizing photons, $\epsilon_{\rm LyC}$, by a factor  
$1+\xi_{\rm LyC}(r)$, in which $\xi_{\rm LyC}(r)$ quantifies the clustering of ionizing sources 
around the central galaxy. The nature of ionzing sources is still unknown, and we assumed 
that $\xi_{\rm LyC}(r)=b_{\rm LyC}b_{\rm LAE}\xi(r)$, in which $\xi(r)$ denotes the 
non-linear matter two-point correlation function, $b_{\rm LAE}$ is the scale dependent bias 
for LAEs from \cite{Ouchi2010} and $b_{\rm LyC}$ is an `effective' bias parameter for the 
ionizing sources. Our fiducial model assumed (for simplicity) that 
$b_{\rm LyC}=b_{\rm LAE}$. We stress that {\it this approach allowed us to quantify our 
ignorance of the population of ionizing sources simply with a single parameter inferred from 
studies of the Ly$\alpha$ forest}, $\epsilon_{\rm LyC}$, and $b^{\rm }_{\rm LyC}$.

To model $f_{c}(r)$, we have taken three different approaches (see \S~\ref{sec:cgm}). 
In the {\it first}, we follow Steidel et al. (2010) who presented a simple 
analytic model for $f_{c}(r)$ which was tuned to reproduce the observed Ly$\alpha$ absorption 
from the CGM (obtained from galaxy-galaxy pairs, see \S~\ref{sec:clumps}). In the {\it second}, 
we include the modification by \cite{DijkstraKramer2012} in the form of a decelerated outflow. For
the {\it third}, we adopt a fitting formula provided by \cite{Rahmati2015} for the {\it area covering 
fraction} of self-shielding gas around simulated galaxies in the EAGLE simulation, which we 
convert into $f_{c}(r)$ via an inverse Abel-transform (see \S~\ref{sec:rahmati}).
 
Our results can be summarized as follows: 
\begin{itemize}[leftmargin=0pt,itemindent=20pt]
\item We find that the `local' ionizing background is boosted by a factor $\sim 50$ 
($\sim 200$ ) at $r=10$ pkpc, where external clustered sources of ionizing radiation dominate 
at $r \gsim $ a few tens ($r \gsim 80-100$) pkpc for the analytical model (two clumpy outflow 
models). These differences arise from the different ionizing photon escape fraction profiles. 
Smaller values for $f_{\rm esc}$ reduce the contribution of the central galaxy to the local 
ionizing radiation field.

\item Our predicted surface brightness profile for the accelerated clumpy outflow model 
falls a factor of $\sim 1.5-2$ below the observations at $r\lsim30$ pkpc and maintains this reasonable 
fit at larger distances. This implies that fluorescence may account for 
50-60\% of the observed flux. The model with the decelerated outflow shows a 
profile above the observations at distances $r\gsim50$ pkpc. The model for the fit to the EAGLE  
simulations predicts a surface brightness which 
is a factor 3-8 below the observations depending on distance (even considering a highly clustered 
population of ionizing sources, $b_{\rm LyC}=3b_{\rm LAE}$).

\item If fluorescence is to reproduce the observed surface brightness profiles, 
then this requires that (\textit{i}) a significant fraction of cold gas, i.e., high covering factor, $f_c$, 
must be present at large distances from the 
center of the galaxy. (\textit{ii}) 
A significant escape fraction of ionizing photons from the central galaxy at small distances is  
required for the central galaxy to reproduce the observations at such distances, and (\textit{iii}) 
the Ly$\alpha$ escape fraction has to be close to 100\%. Otherwise, fluorescence only 
accounts for a fraction of the flux in Ly$\alpha$ halos.

\end{itemize}

There are significant uncertainties associated with predicting the fluorescent Ly$\alpha$ flux in halos surrounding star forming galaxies, including the unknown nature of sources of ionizing radiation and the gaseous content of the CGM. These uncertainties represent key uncertainties in our understanding of galaxy formation and evolution, and in how ionizing (and Ly$\alpha$) photons escape from the interstellar media of galaxies. Our calculations indicate that a contribution from fluorescence 
to the total flux in the halo is possible, and more if ionizing sources and self-shielded gas are correlated. This correlation may help to reduce the contribution to the fluorescent Ly$\alpha$ surface brightness from the central galaxy at $r\sim$ 30-70 pkpc. If the central galaxy was entirely responsible for the fluorescent signal at these impact parameters, then this may make it difficult to the environmental dependence of the scale-length of LAHs \citep[and the non-dependence on $M_{\rm UV}$ of the central galaxy, ][]{Steidel2011, Matsuda2012,Momose2015,Wisotzki2015}.

In spite of theoretical uncertainties, we expect rapid progress in this field due to anticipated improvements in the quality of incoming data. As our previous discussion illustrates, comparing the Ly$\alpha$ and UV continuum surface brightness profiles provides useful constraints on the models. Another very useful observable is the spectrum. With integral field spectrographs such as MUSE\footnote{\url{http://www.eso.org/sci/facilities/develop/instruments/muse.html}} it is possible to constrain the spatially resolved spectrum. As alluded to in \S~\ref{sec:intro}, the Ly$\alpha$ spectral line shape contains information on the kinematics of the scattering medium. Different models for Ly$\alpha$ halos generally  predict different Ly$\alpha$ spectra, and how they vary with position. Spectra of Ly$\alpha$ created as fluorescence have been predicted to be narrow, and double peaked \citep[e.g.,][]{GouldWeinberg1996,Cantalupo2005}. Further into the future, it would be extremely valuable to use, e.g., JWST\footnote{\url{http://jwst.nasa.gov/index.html}} and/or future ground-based facilities such as E-ELT\footnote{\url{https://www.eso.org/sci/facilities/eelt/}} or GMT\footnote{\url{http://www.gmto.org/}} 
to search for spatially extended H$\alpha$ emission. H$\alpha$ emission is also produced by fluorescence, but does not resonantly scatter through neutral hydrogen gas. Future joint H$\alpha$ and Ly$\alpha$ observations of halos would provide independent constraints on all these halos. 
Independently, measurements of the polarization of Ly-alpha would differentiate
between models that invoke scattering and in-situ production to explain the observed
spatially extended Ly-alpha halos \citep[see, e.g.,][]{DijkstraLoeb2008,DijkstraKramer2012}.
First detections of polarization in spatially extended Ly$\alpha$ emission have recently
been reported \citep[][and also see \cite{Prescott2011} for a non-detection]{Hayes2011,Humphrey2013,Beck2016}, which provide independent constraints on the models.

While none of these measurements are easy, and their interpretation complicated, it is important to understand Ly$\alpha$ halos, as they contain information on the gaseous content of the CGM which complements that of absorption line studies \citep[e.g.,][]{ DijkstraKramer2012, HennawiProchaska2013}, and on the amplitude of the local ionizing radiation field. The latter contains information on quantities like the escape fraction, and on the nature of ionizing sources in general. Both of these questions represent some of the main uncertainties in our understanding of the reionization process, and the physical properties of the galaxies in general.

\acknowledgments
We are grateful to the anonymous referee, whose constructive comments 
allowed for a significant improvement of our work. We thank Alireza Rahmati 
for comments about the Rahmati et al. model.
We acknowledge Stuart Wyithe and Avi Loeb for their opinions and ideas.
We thank Max Gronke for useful discussions about the clumpy medium
and the importance of the escape fraction around and within galaxies; we
also thank Brendan Griffen, Maxime Trebitsch and J.Xavier Prochaska  for their 
advice in clustering aspects. We thank the astronomy department at Columbia 
University and at UC Santa Barbara for their kind hospitality.


\appendix
\section{Parameters of the clumpy circum-galactic medium}\label{sec:clumpfig}

\begin{figure}[h]
\includegraphics[width=1 \textwidth]{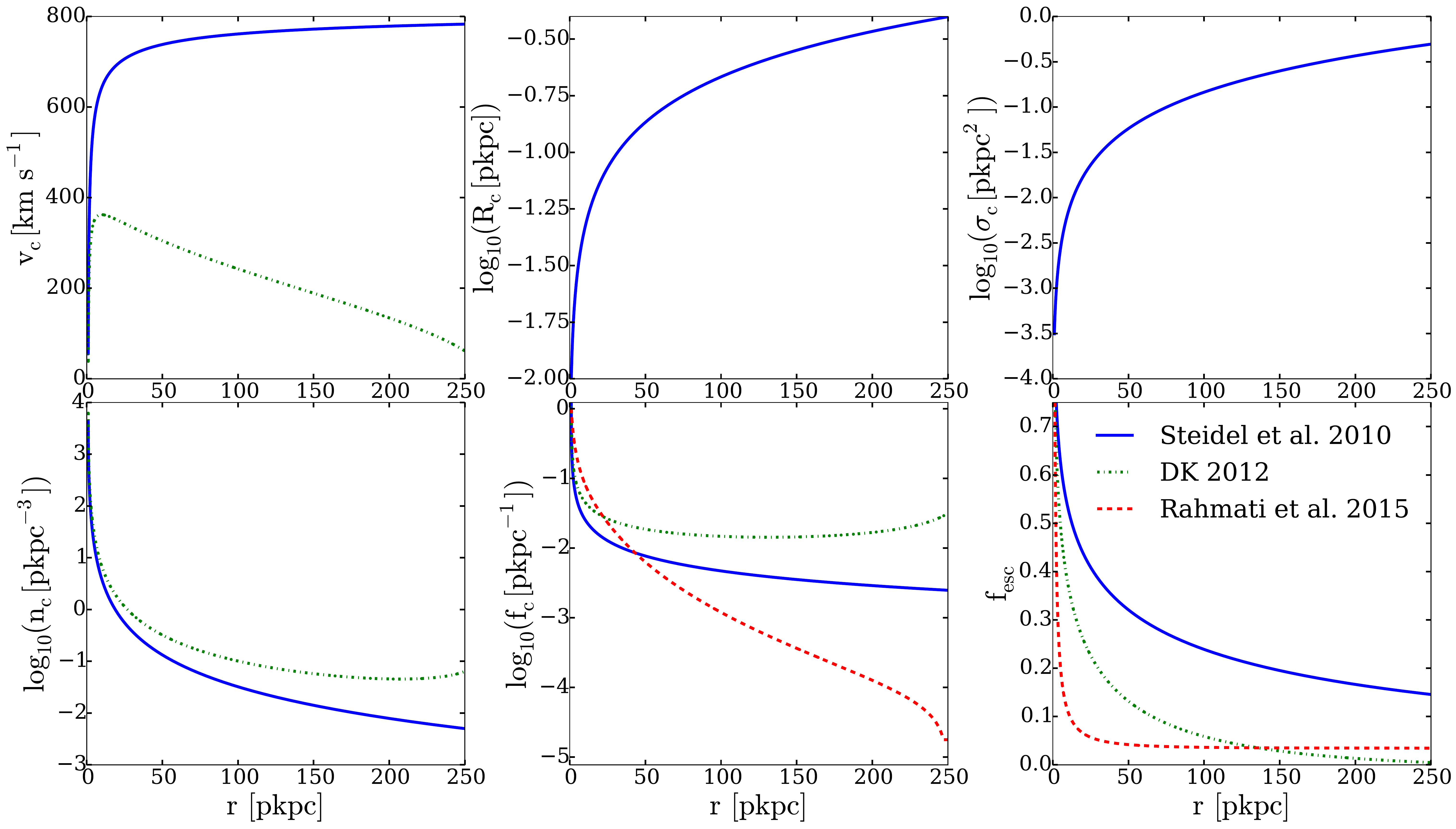}
\caption{Parameters for the clumpy outflow model with respect to distance from the 
central LAE. {\it Blue solid line} denotes the Steidel et al. method  
and the {\it green dot-dashed line} that of Dijkstra \& Kramer. The covering factor and escape fraction 
computed from the results by Rahmati et al. are also shown as {\it dashed red line}.
Note the large differences for the case of the covering factor and escape fraction using the 
Rahmati et al. formalism compared to the other two outflowing models. See the text above for the 
description of the parameters.}
\label{fig:clumps}
\end{figure}

{\it Blue solid lines} in Figure \ref{fig:clumps} show the parameters of the clumpy 
outflow model using the method of \cite{Steidel2010}. The decelerated outflow model by 
\cite{DijkstraKramer2012} (their model IV) is represented by the {\it dot-dashed green lines}  
where the two models
present differences. {\it Dashed red lines} denote the parameters computed from the results  
by \cite{Rahmati2015}. From left to right and top to bottom, we present the outflow 
velocity $v_{c}(r)$, the radius of the clumps ${\rm R_c}(r)$, the area of the clumps
$\sigma_{c}(r)$, the number density of clumps $n_c(r)$, the covering factor $f_c(r)$
and the escape fraction of ionizing photons $f_{\rm esc}(r)$. All these values are shown 
with respect to the distance from the central LAE. 
We can see that the covering factor for the case of the Rahmati et al. model presents a much 
steeper decay at large distances compared to the other two models, although it reaches 
higher values at small distances. This enhances the number of clumps close to the centre 
and reduces it at large distances, which in turn, makes the escape fraction 
to suffer a very strong decrease within a few tens of kpc. At the maximum distance of our 
calculations, $r=250\,{\rm pkpc}$, the three methods reach an escape fraction value consistent 
with other works; The Steidel et al. model yields a value around $15\%$, Rahmati et al. around 
$5\%$ and Dijkstra \& Kramer model around $0\%$. As a reference, the work by \cite{Kuhlen2012} 
finds a value for the average escape fraction $\sim3\%$ at $z=3$ (Table \ref{ta:params}). 
This value is inferred from studies of the Ly$\alpha$ forest although large differences are 
obtained between works concerning galaxy surveys.

\section{Non-linear clustering of LAEs}\label{sec:cluster}

Clustering of LAEs is commonly addressed making use of a power
law of the form 
\begin{equation}\label{eq:xi}
\xi(r)= \left(\frac{r}{r_0}\right)^{\alpha_c}~,
\end{equation}
with values $r_0\sim2.5\,{\rm Mpc\,h^{-1}}$ (this parameter depending on redshift) 
and $\alpha_c\sim-1.8$ \citep[e.g.,][]{Bielby2015, Gawiser2007, Kovac2007, 
Ouchi2003, Ouchi2010, Guaita2010}. However, for the purpose of our work, 
we are interested in the effect due to galaxies at small distances from 
the central LAE. This cannot be taken into account using a power law but 
a proper non-linear clustering treatment is necessary \citep[][]{Iliev2003, Sheth2001b}. 
In order to address this more realistic clustering analysis, we make use of the 
\textit{halo model} \citep[see, e.g.,][and also \cite{Cooray2002} for an extensive description]
{Zheng2015, Sheth2001,Sheth2001b, Smith2003, Peacock1996}, which has been largely
used in order to reproduce simulations and observations providing satisfactory
results \cite[e.g.,][]{Genel2014, Tal2013, Sales2007, Nagai2005}.
We construct an initial matter power spectrum \footnote{We consider that the 
dark matter power spectrum is the same as the matter power spectrum, since the 
former is the major contributor to the latter.} 
\citep[see, e.g.,][for a review on the calculation of the linear matter power spectrum]
{Peacock1994} and use the publicly available software \textit{hmf}
\footnote{\url{https://github.com/steven-murray/hmf}} to compute the non-linear matter power 
spectrum at the redshift of interest. Once we have the power spectrum, we compute the 
non-linear matter two-point correlation function as 
$\xi(r)=\xi(r)_{\rm 1h}+\xi(r)_{\rm 2h}$. Finally, we make a simple fit to the data of \cite{Ouchi2010}
to obtain the scale dependent bias of LAEs. Below, we describe
this two step procedure in more detail.

\subsection{Non-linear matter power spectrum and correlation function}\label{sec:corr}

\begin{figure*}[h]
\includegraphics[width=0.5 \textwidth]{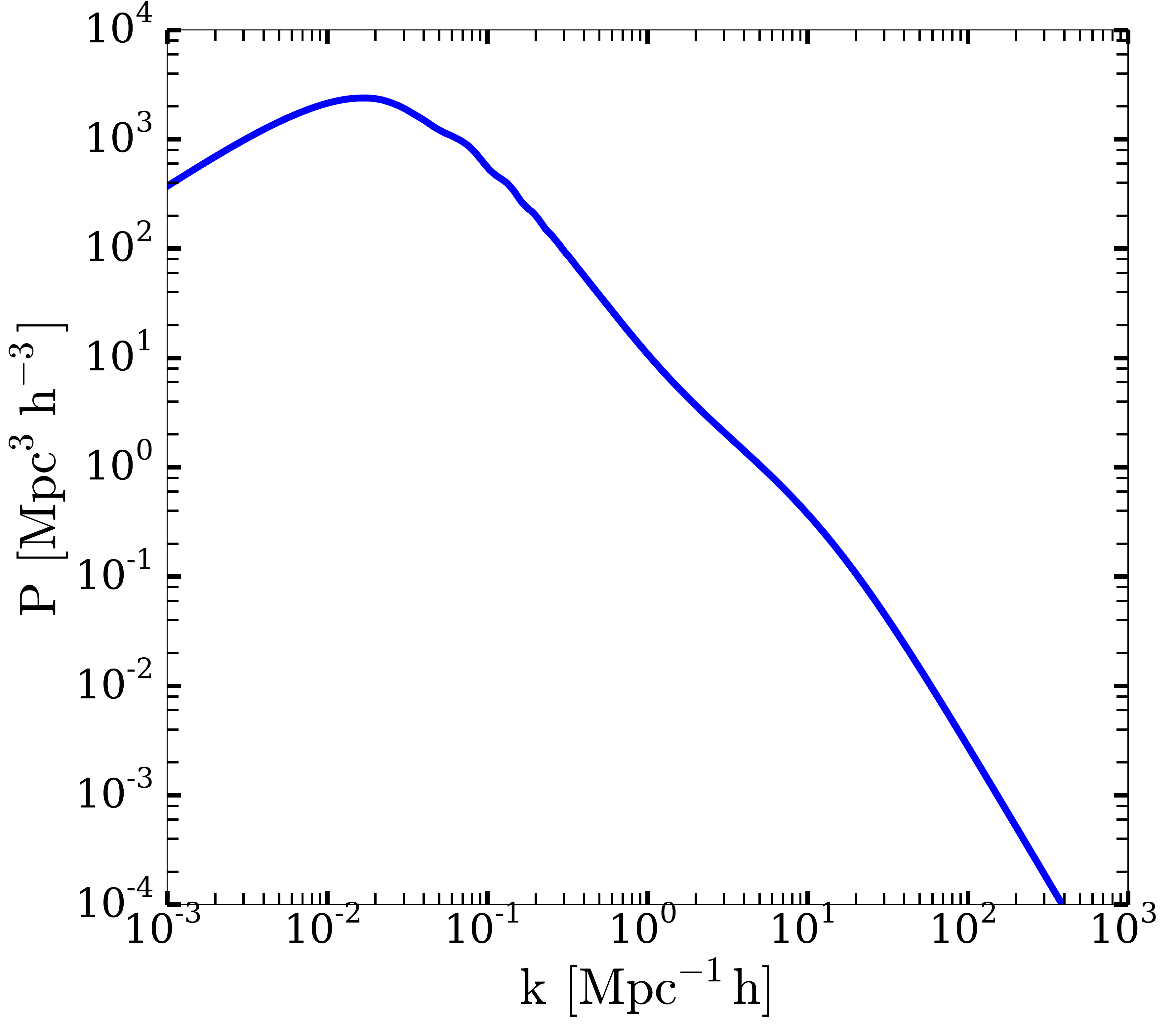}\includegraphics[width=0.48 \textwidth]{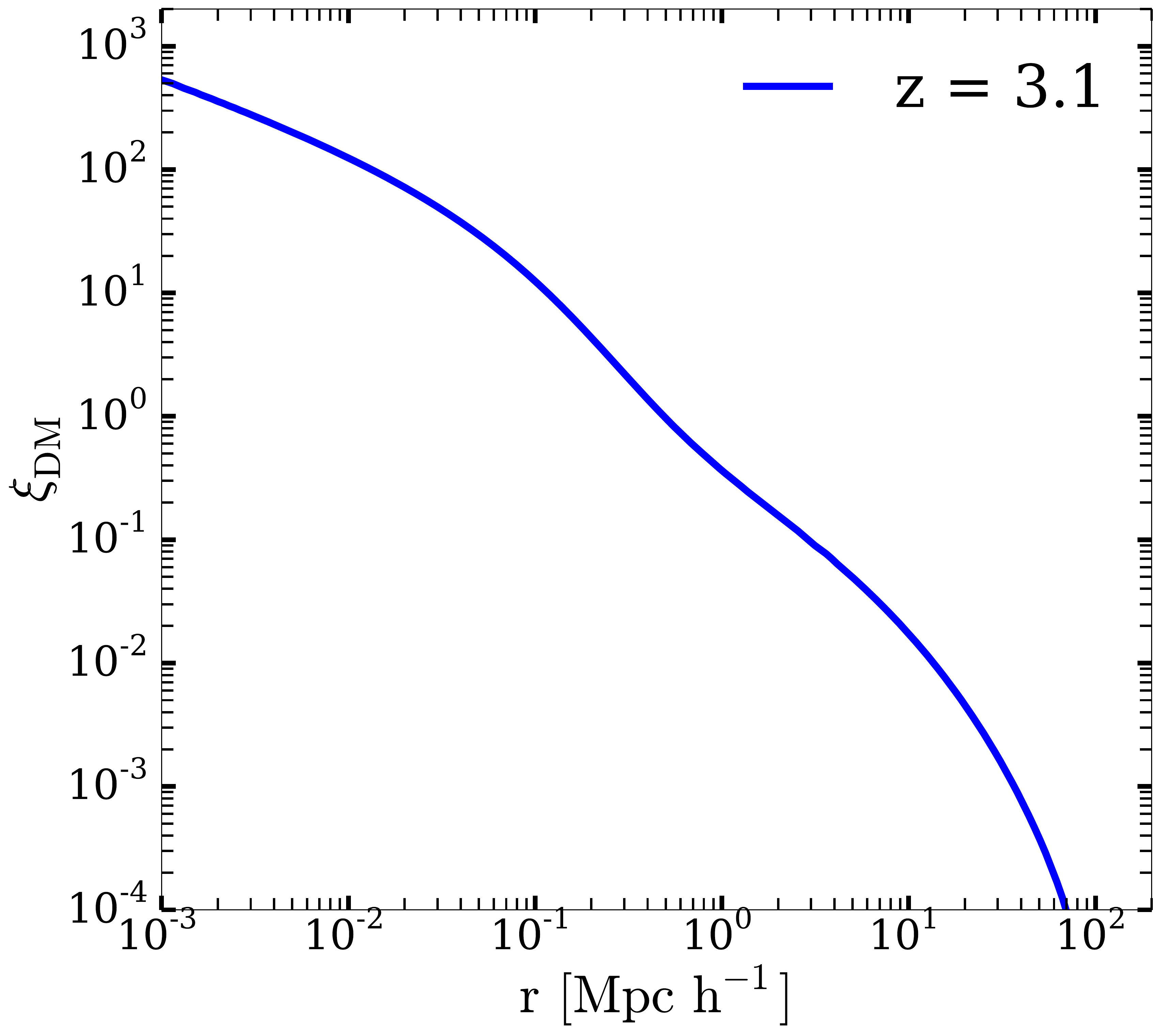}
\caption{\textit{Left panel}: The matter power spectrum computed at  
redshift $z=3.1$. Notice the effect of the non-linear clustering between 
$k\sim1\,{\rm Mpc^{-1}\,h}$ and $k\sim10\,{\rm Mpc^{-1}\,h}$ seen as a change in the slope 
which produces an enhancement of the power at small scales. 
\textit{Right panel}: The two-point correlation function for dark matter. 
The 2-halo term dominates at distances larger than $r\sim1\,{\rm Mpc\,h^{-1}}$ and 
denotes the linear clustering term in the correlation. The 1-halo term, denoting the 
clustering of sources inside the same dark matter halo, dominates at shorter scales 
and is the responsible for the enhanced non-linear part of the correlation function.}
\label{fig:pow}
\end{figure*}

We compute the non-linear matter power spectrum using \textit{hmf}, adopting the default 
settings except for the redshift at which the calculation is performed. 
\textit{hmf} makes use of the software CAMB \citep{Lewis1999} to compute the 
transfer function, and applies the non-linear corrections to the power spectrum 
using the HALOFIT \citep{Smith2003} model, with the updated parameters from 
\cite{Takahashi2012} (see a detailed description of the method in \url{https://github.com/steven-murray/hmf}).
For these calculations, a $\Lambda$CDM cosmological model with parameters
$\Omega_{\Lambda}=0.6825$, $\Omega_b=0.049$, ${\rm h=H_0/100=0.67}$ and 
$\sigma_8=0.8344$ is assumed. 
Left panel in Figure \ref{fig:pow} shows the computed matter power spectrum, 
where the effect of the non-linear clustering can be observed as a 
change in the slope between 
$k\sim1\,{\rm Mpc^{-1}\,h}$ and $k\sim10\,{\rm Mpc^{-1}\,h}$, thus enhancing the power 
at small scales, i.e., large $k$. 
Once the power spectrum is computed, the correlation function can be easily 
obtained using the Fourier transform of the former as
\begin{equation}\label{eq:corr}
\xi_{DM}(r)=\frac{1}{4\pi^2}\int{\rm d}k\,k^2P(k)\frac{\sin(kr)}{kr} ~,
\end{equation}
where $k$ is the wavenumber. The right panel in Figure \ref{fig:pow} shows the 
correlation function for dark matter. The contribution of the two halo terms is clearly 
visible: at distances larger than $r\sim1\,{\rm Mpc h^{-1}}$ the 2-halo term dominates, 
denoting the linear part of the correlation. At shorter distances, the 1-halo term is the 
responsible for the non-linear effect that raises the value of the correlation function at 
those scales.

\subsection{Bias and LAE correlation function}\label{sec:bias}
\begin{figure*}[h]
\includegraphics[width=0.5 \textwidth]{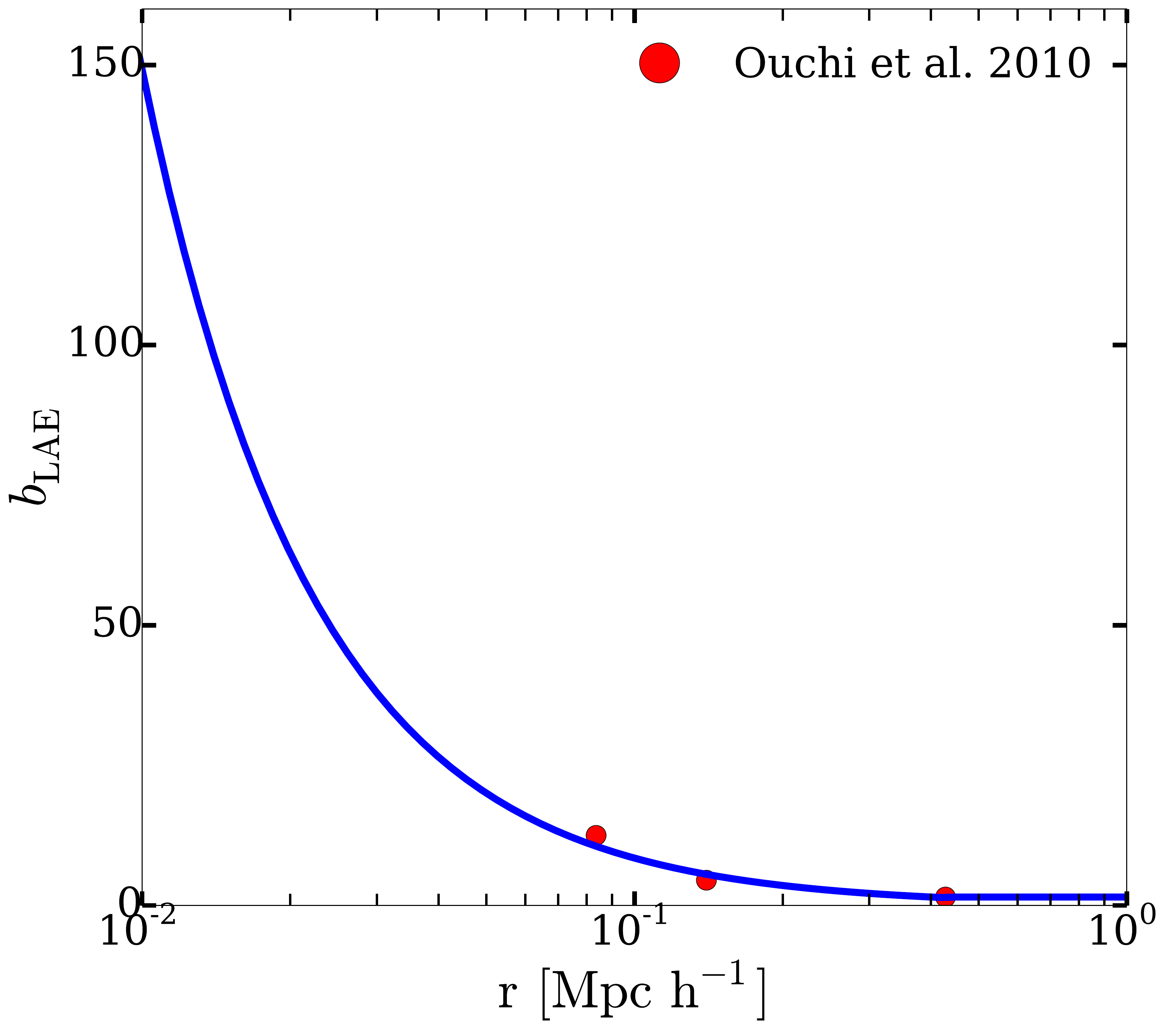}\includegraphics[width=0.48 \textwidth]{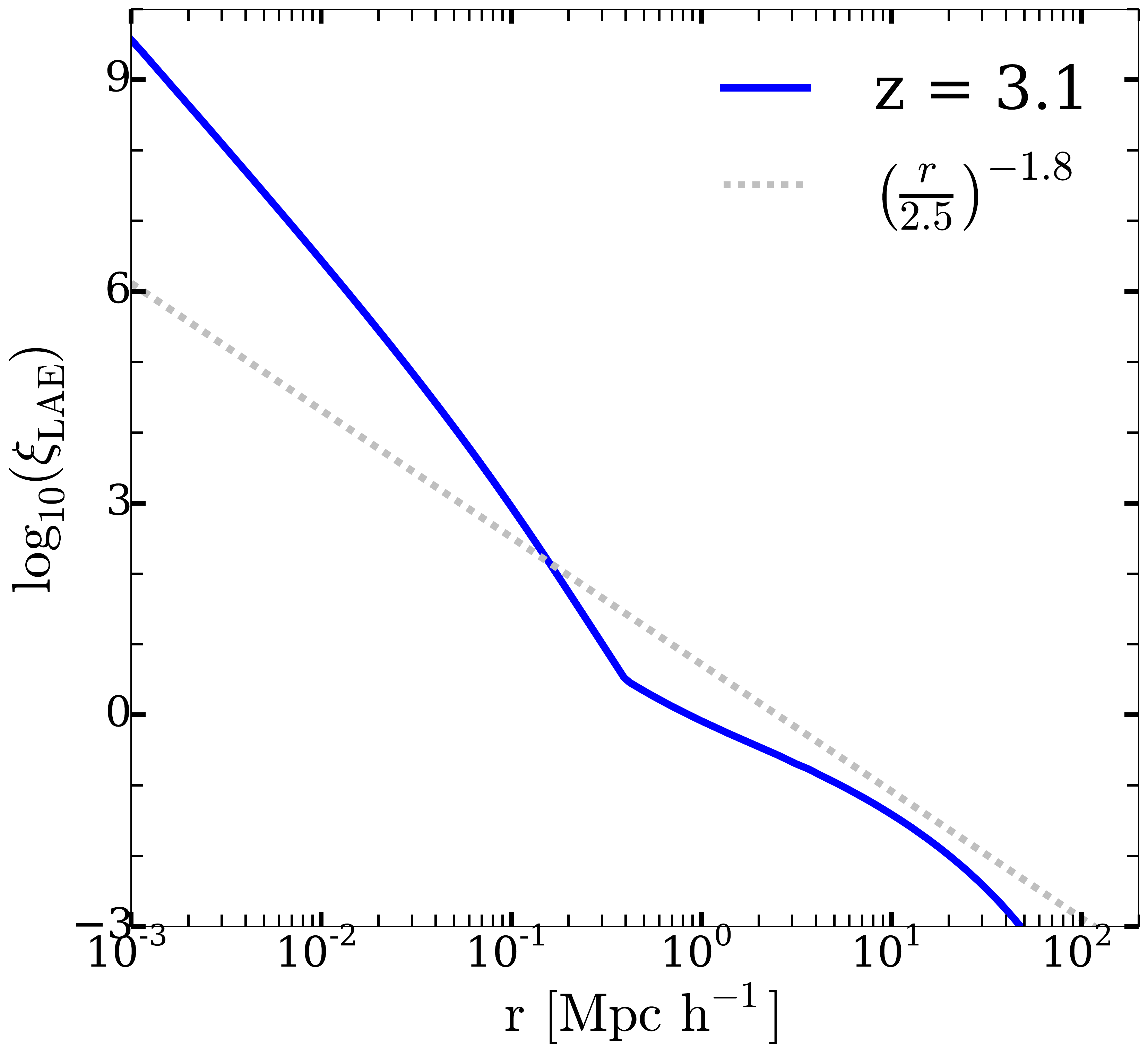}
\caption{\textit{Left panel}: The distance dependent bias for the case of LAEs. The {\it red dots} 
denote the data points from \cite{Ouchi2010} used for fitting the bias profile. \textit{Right panel}: 
The {\it blue solid line} denotes the correlation function for LAEs. For comparison,
the {\it grey dashed line} represents a commonly adopted correlation function with  
correlation length $r_{0}=2.5\,{\rm Mpc\,h^{-1}}$ and power law index $\alpha=-1.8$.}
\label{fig:bias}
\end{figure*}

The final step for the calculation of the correlation function of LAEs is
to consider the bias between LAEs and dark matter, which is described as
\begin{equation}
b^2_{\rm LAE}(r)=\frac{\xi_{\rm LAE}(r)}{\xi_{\rm DM}(r)} ~.
\end{equation}
We use the data in 
\cite{Ouchi2010} who provides the scale dependent bias for LAEs at several redshifts. 
The presence of faint sources below the observability 
threshold or the resolution power makes the observation of clustering very difficult 
\citep{Gawiser2007,Kovac2007, Ouchi2010}. Due to this, we find only three data points at 
distances $r<0.4\,{\rm Mpc\,h^{-1}}$ \citep[see lower left panel in Figure 11 in][]{Ouchi2010}.
We perform a linear fit to the data points, in log-log representation, and assume the 
constant average bias computed by \cite{Ouchi2010}, $b_{\rm LAE}\sim1.5$, for larger distances. 
We note that the resulting correlation function of LAEs at small scales is very sensitive to the value 
of the distance dependence bias. However, an unavailable larger number of data points would be 
required to obtain a more reliable fit at such distances. The left panel in Figure \ref{fig:bias} shows 
the bias profile and the data from \cite{Ouchi2010} and the right panel the resulting LAE 
correlation function.  

\section{Effects of varying the galaxy bias}\label{sec:rah}
We show here the surface brightness profiles obtained when allowing small 
variations of the galaxy bias. We address two cases: (\textit{i}) We consider the effect of 
a higher LAE overdensity since in our work we have been comparing to the highest density 
case from \cite{Matsuda2012}. We do this simply by doubling the default value of $b_{\rm LAE}$. 
We call this bias $b_{\rm LAE}^{2x}$ and its total effect is an increase a factor 4 for the 
correlation function. We adopt this simple procedure because we ignore how the 
clustering profile of a `more' overdense region may change compared to the computed one.
The overdensity value is now $\delta_{\rm LAE}\sim5.8$, slightly above 
the maximum value of \cite{Matsuda2012}. (\textit{ii}) We also consider a departure from 
the LAE bias when considering the bias of ionizing sources. We assess the value 
$b_{\rm LyC}(r)=3\times b_{\rm LAE}(r)$. Larger values appear to be not physically motivated
to us. In this case, the total effect is a boost a factor 3 for the correlation function, so we 
expect to find very similar results in the two cases.

\begin{figure*}[h]
\includegraphics[width=0.34 \textwidth]{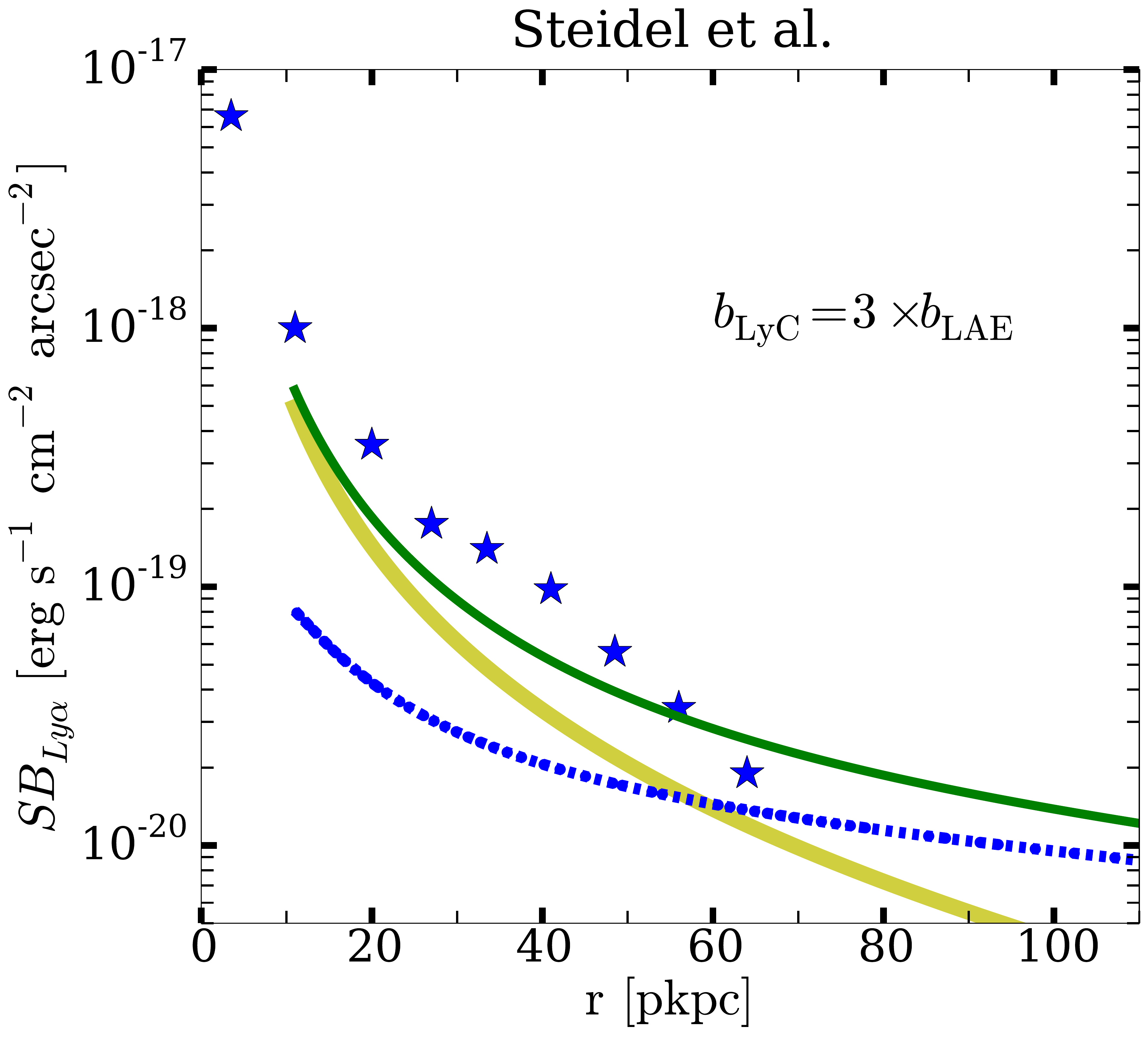}\includegraphics[width=0.32 \textwidth]{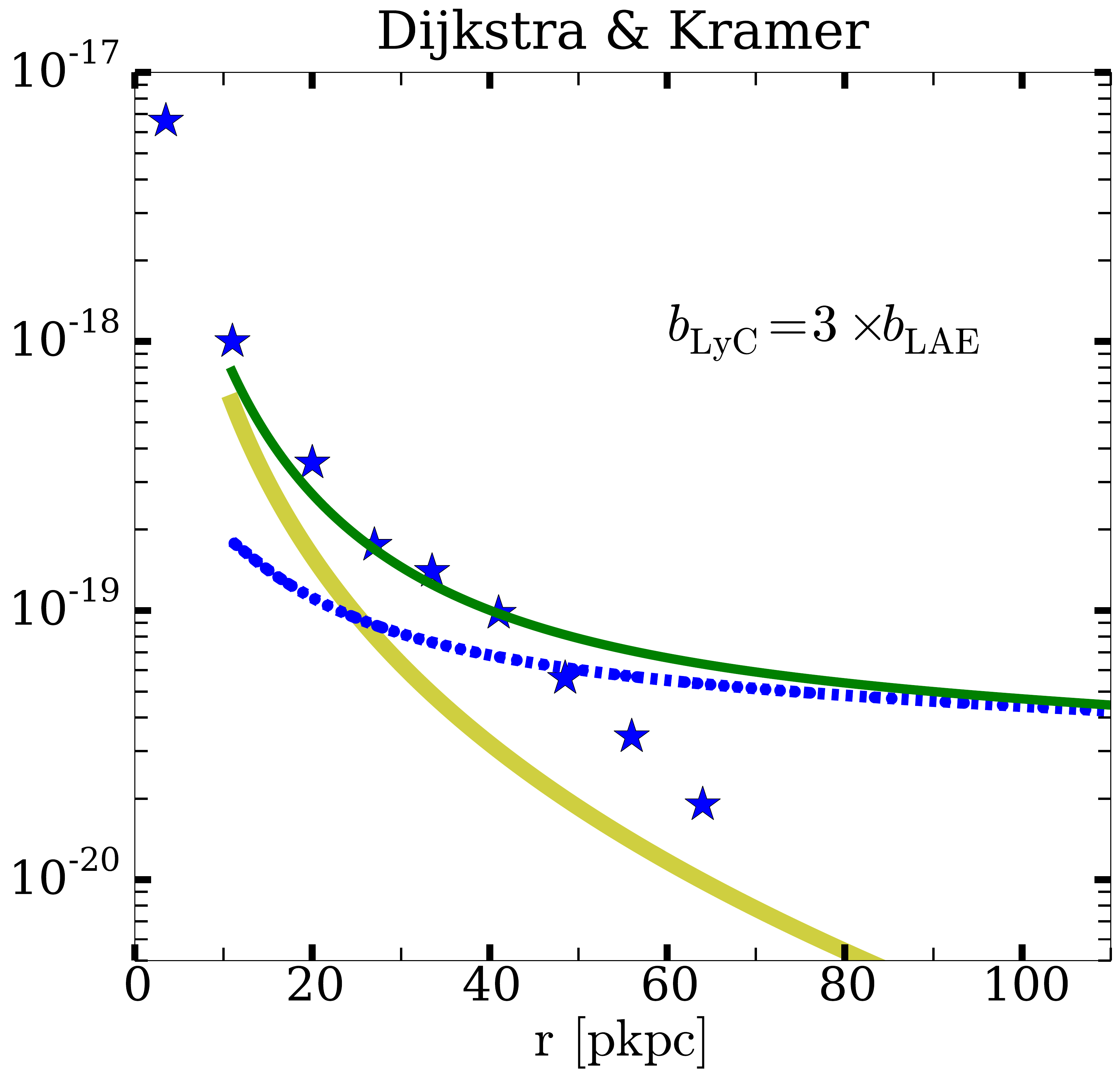}\includegraphics[width=0.32 \textwidth]{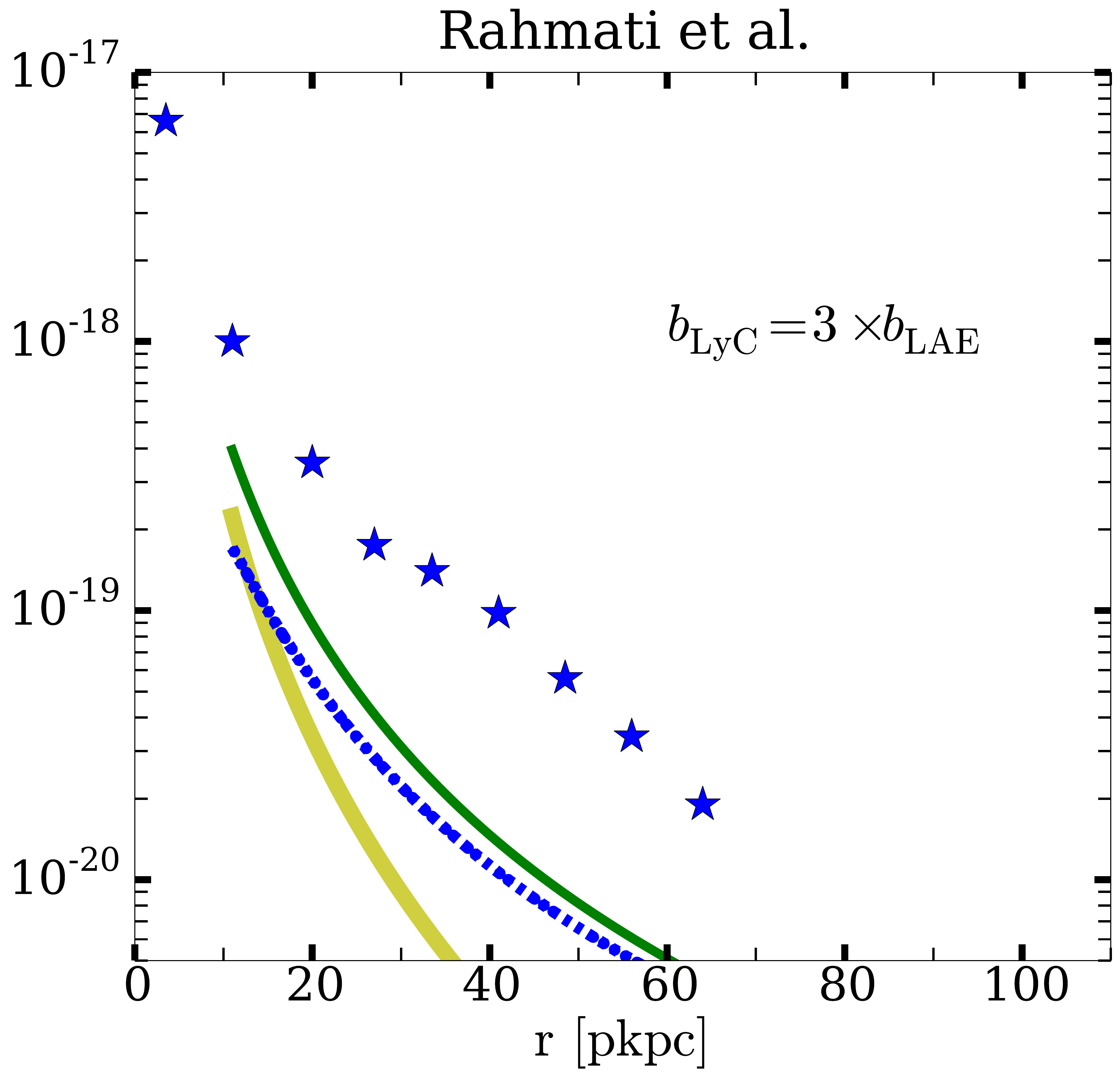}
\includegraphics[width=0.34 \textwidth]{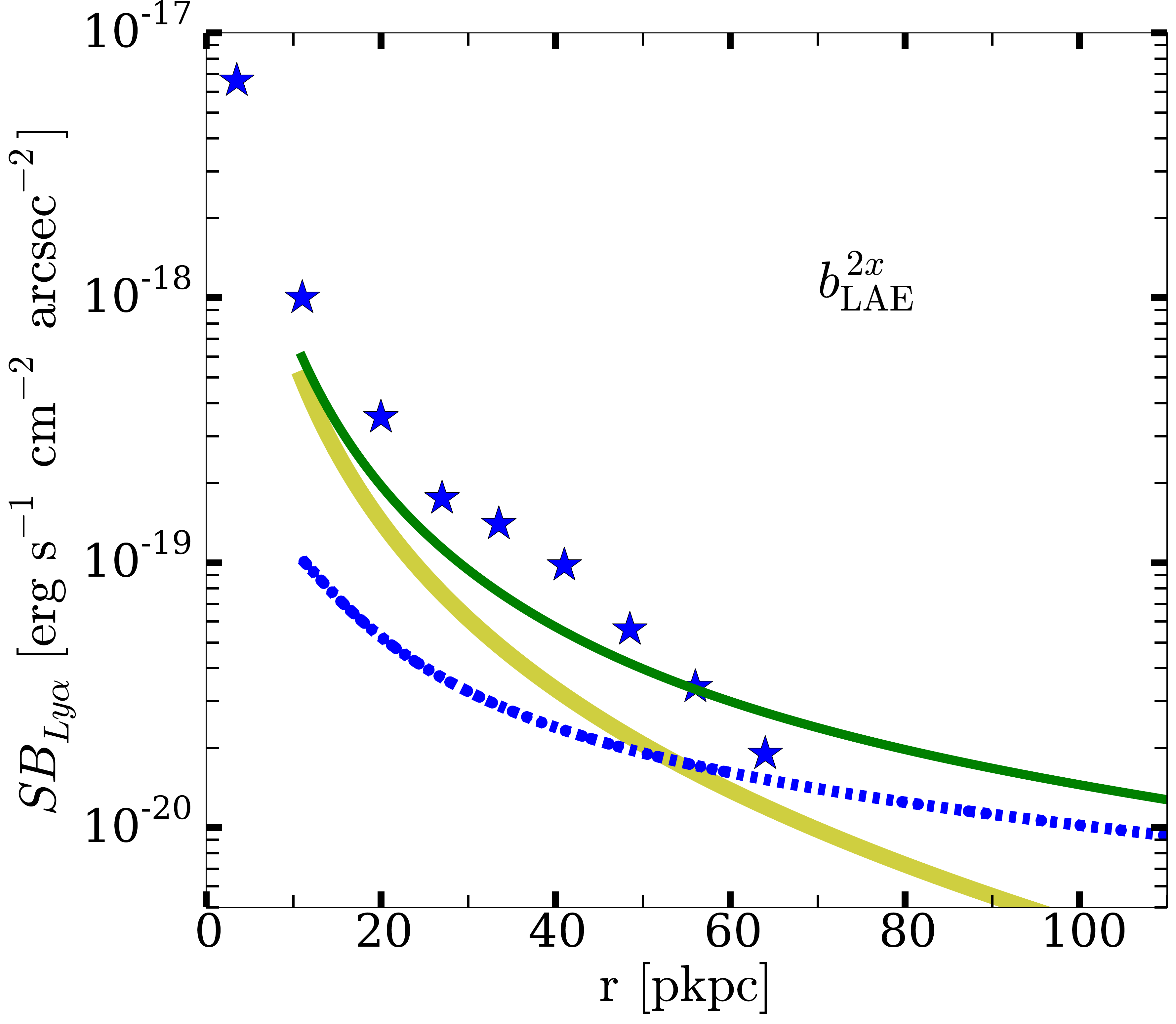}\includegraphics[width=0.32 \textwidth]{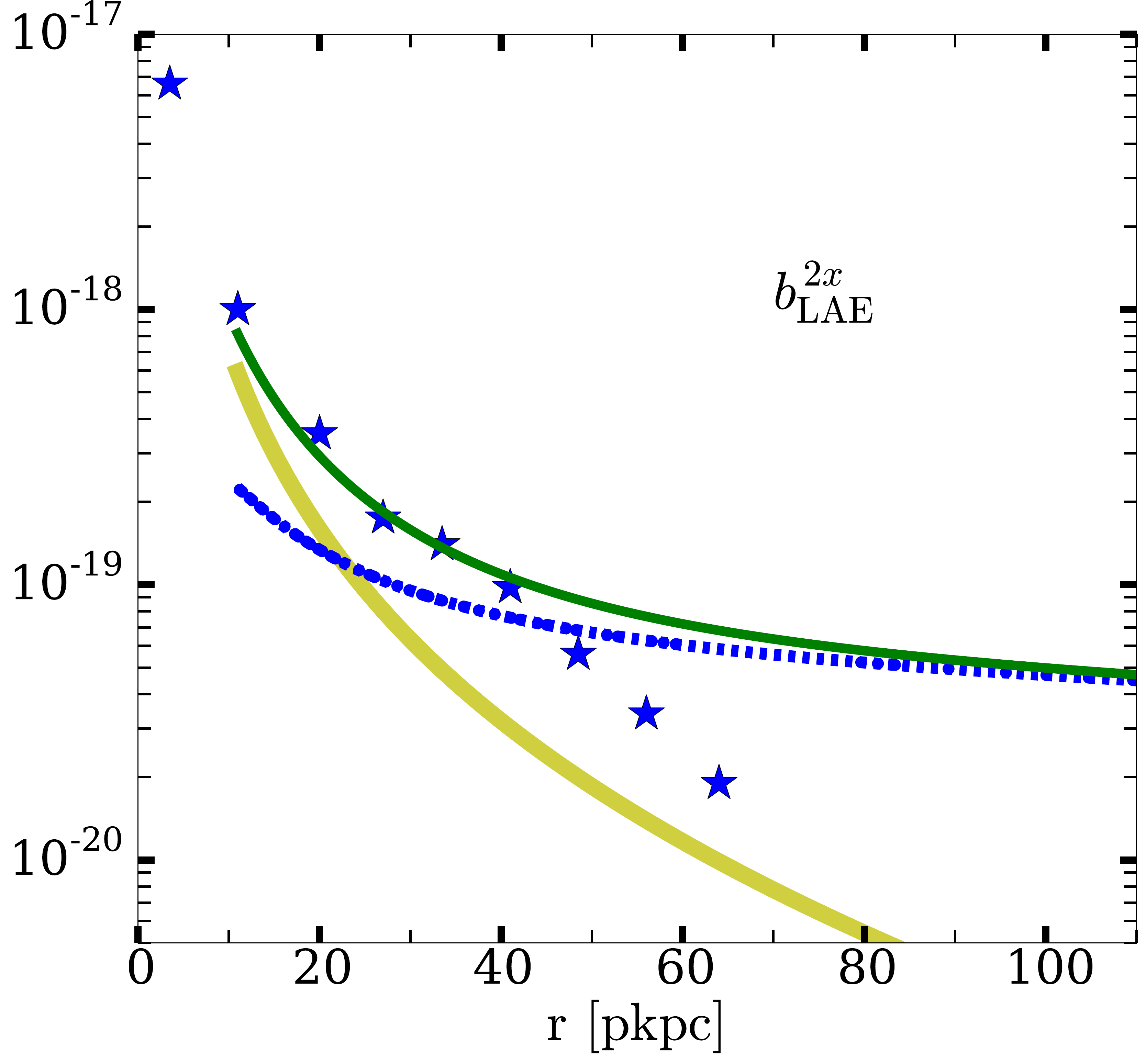}\includegraphics[width=0.32 \textwidth]{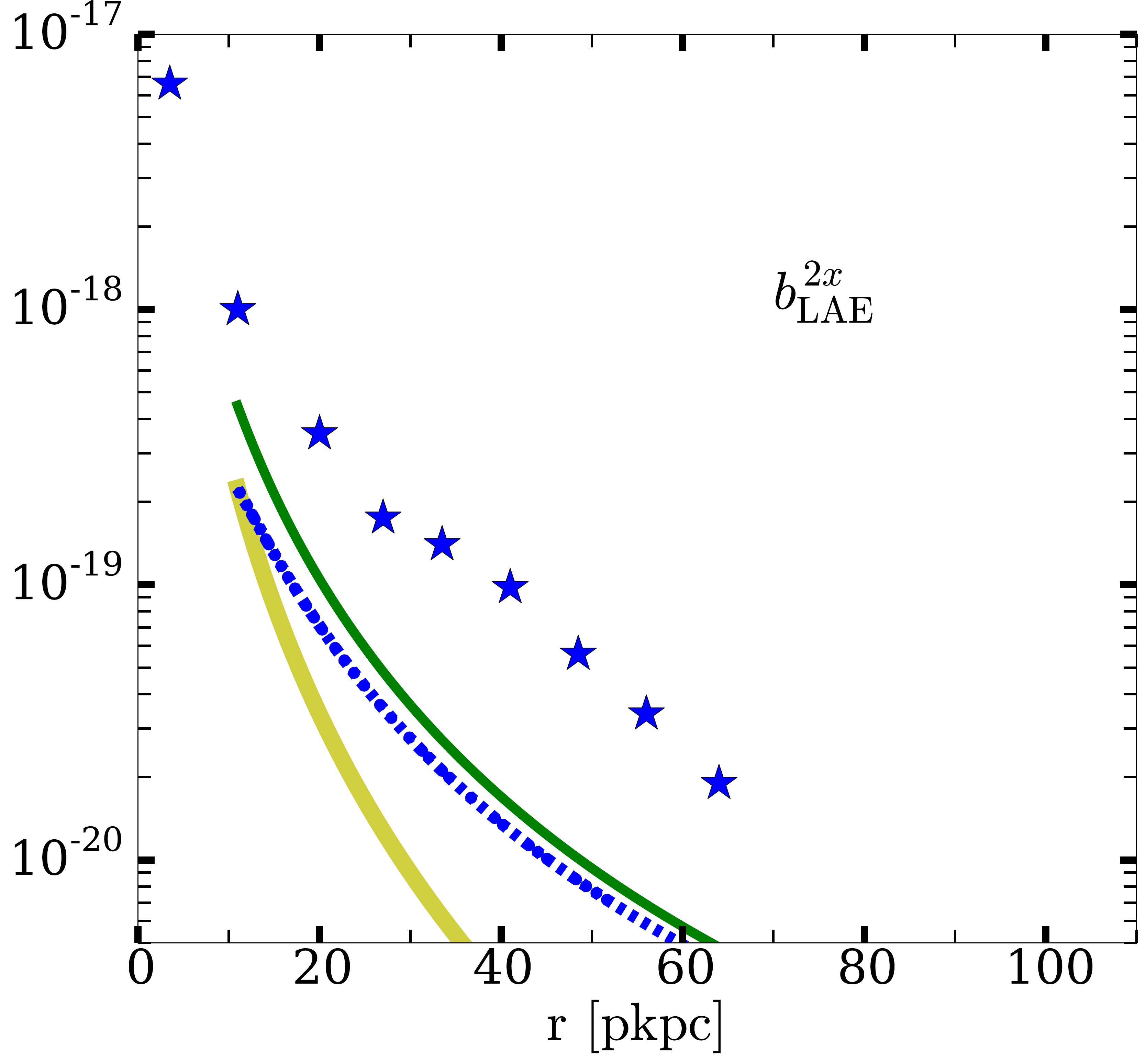}
\caption{\textit{Upper panels} show the surface brightness profile when accounting for a 
different bias between LAEs and ionizing sources. \textit{Lower panels} show the profiles  
when doubling the LAE bias in order to recover higher overdensities. The markers and lines 
are the same as in Figure \ref{fig:sbz3}. 
Due to the 
similar total effect, there are very small differences between the two cases. There are also 
small differences with our default models which do not change our main results.}
\label{fig:rah}
\end{figure*}

Figure \ref{fig:rah} shows the results for these two calculations. Due to the similar 
total effect to the correlation function, there are no significant differences between the two
methods. More important, there are no significant changes when comparing to our default 
model parameters. Thus, our main conclusions are not very sensitive to small variations 
of the bias parameter but they are strongly dependent on the properties of the medium.

\newpage
\bibliographystyle{apj}
\bibliography{lyahaloapj}\label{References}


\end{document}